\documentclass[fleqn,12pt]{wlscirep}
\usepackage[utf8]{inputenc}
\usepackage[pagewise]{lineno}
\usepackage{rotating}
\usepackage[T1]{fontenc}
\usepackage{pdflscape}
\def\bm#1{{\mbox{\boldmath{$#1$}}}}
\usepackage{hyperref}
\usepackage{float}
\usepackage{graphicx}
\usepackage[section]{placeins}
\usepackage{setspace}
\usepackage{adjustbox}
\usepackage{lineno}
\usepackage{lscape}
\usepackage{bm}
\usepackage{mathtools}
\usepackage[superscript]{cite} 
\usepackage{subcaption}
\usepackage{amsmath}
\usepackage{float}
\usepackage{tabularx}
\usepackage{todonotes}
\usepackage{booktabs}
\usepackage{multirow}
\usepackage[table]{xcolor}
\usepackage{array}
\usepackage{booktabs}
\usepackage{multirow}
\usepackage{microtype}
\usepackage{graphicx}

\title{A Dynamic Joint Latent Class Model For Longitudinal and Survival Data}

\author[1]{Q. Selimi}
\author[1,$\dagger$]{C. Charalambous}
\author[1, $\dagger$]{T. Baghfalaki}

\affil[1]{Department of Mathematics, The University of Manchester, Manchester, UK}
\affil[*]{Corresponding author: T. Baghfalaki: taban.baghfalaki@manchester.ac.uk}
\affil[$\dagger$]{these authors contributed equally to this work}

\begin{abstract}
{\footnotesize In this paper, we propose a new Dynamic Joint Latent Class Model (DJLCM)  by extending the traditional JLCM to incorporate a time-varying class membership submodel. We use a first-order discrete time Hidden Markov Model to govern covariate-dependent transitions. The longitudinal outcome, given class membership at each time point, is modelled using a linear mixed-effects model. Additionally, we include a survival submodel conditional on the class membership, capturing the association between longitudinal trajectories and time-to-event outcomes.
 Instead of the traditional likelihood function, we employ the forward algorithm to account for the dynamic nature of the latent classes and provide computationally efficient inference.
For parameter estimation, we adopt a Bayesian framework using Hamiltonian Monte Carlo (HMC).
We evaluated the proposed model through extensive simulation studies which demonstrated good convergence diagnostics and excellent discriminative power of our model. We applied the DJLCM to the colorectal cancer dataset, analysing the association between repeated tumour size measurements and time to death in patients with metastatic colorectal cancer. Our findings indicate that KRAS mutation status and treatment arm did not significantly affect the tumour size or event risk in either class. Finally, we compared the DJLCM to two JLCMs in the literature, demonstrating its superior performance in terms of model selection criteria and the insights it provides.
}
\end{abstract}

\begin{document}
\flushbottom
\maketitle
\thispagestyle{empty}

\section{Introduction}

In many oncology studies, repeated measurements of prognostic markers and time to a clinical event of interest are recorded, aiming to monitor disease progression in patients with cancer. As these two clinical outcomes tend to be correlated when measured on the same subjects, modelling them separately may lead to biased estimates \cite{guo2004separate}. Indeed, common survival models used in the literature, such as the Cox proportional hazards model \cite{cox1972regression}, fail to fully incorporate the disease information contained in the longitudinal biomarker. As an alternative, two-stage approaches were introduced \cite{albert2010estimating}, yet they underestimate the association between the longitudinal and survival process, as the first stage does not account for the information carried by the event process. 
\\
To address this, joint models for longitudinal and survival data were developed, which simultaneously model the longitudinal and survival outcomes by considering the joint distribution of the two processes. 
The most common joint models in the literature are shared random effects models, where the longitudinal and survival outcomes are linked together through subject-specific shared random effects \cite{faucett_and_thomas_1996, tsiatis1995modeling}. The longitudinal biomarker trajectory is typically modelled as a linear mixed-effects model and the hazard of the event commonly modelled via a Cox proportional hazards model. The random effects from the longitudinal trajectory are incorporated in the survival model to account for the association between the two processes. 
To this day, shared random effects models have been extended in several aspects: to accommodate non-normal longitudinal outcomes \cite{faucett1998analysis, brilleman2016joint, li2010joint,ganjali2026unified,ganjali2024joint,baghfalaki2014bayesian}, multivariate/mixed longitudinal biomarkers \cite{chi2006joint, long2018joint, papageorgiou2019overview, rizopoulos2011bayesian,baghfalaki2014joint,malekpour2026joint,baghfalaki2026two}, competing risks and multi-state models for the survival process \cite{ferrer2016joint, andrinopoulou2014joint, lu2017bayesian, huang2011general,arabi2024bayesian,mehdizadeh2021two,baghfalaki2020bayesian}, and different  association structures \cite{mauff2017extension}. 
\\
However, the shared random effects model assumes that differences in trajectories and event risk are entirely explained by the individual-level random effects. In some situations, there exist underlying distinct profiles of trajectory and event risk evolution within the population. To account for this, the joint latent class model (JLCM) \cite{lin2002latent, proust2014joint} was developed. The JLCM assumes that the population consists of several subpopulations, with different trajectory features and event risk profiles, allowing for identification of patients with distinct disease mechanisms. The JLCM consists of a class membership model, where the class membership probability typically follows a multinomial logistic regression model, and class-specific mixed-effects and Cox proportional hazard models for the longitudinal and event outcome, respectively. 
The joint latent class model has been extended to allow for recurrent events \cite{han2007parametric}, multivariate or non-gaussian longitudinal outcomes \cite{taylor2006nonlinear, proust2009joint}, and competing risks \cite{proust2013analysis}. The standard JLCM assumes that the longitudinal and survival outcomes are linked through the latent classes, conditional on which, the two processes are independent. This assumption implies that the longitudinal trajectory affects the risk of event only through the classes. To assess the validity of this assumption, Jacqmin Gadda~\cite{jacqmin2010score} developed a score test for conditional independence in JLCMs.
\\
To relax this assumption and allow individual differences within classes to affect the hazard, Garre et al.\cite{garre2008joint} proposed a Bayesian joint latent class change-point model in which the longitudinal creatinine measurements and the risk of graft failure are linked together through both the latent classes and the shared random effects, conditional on which, the two processes are independent. This model can be considered as a shared parameter within joint latent class model since it links the longitudinal and survival process through the random effects, while simultaneously allowing for latent subgroups in the population. However, the authors only considered a fixed number of classes. Motivated by a study on patients with cystic fibrosis where the time to first aggravation was of interest, Andrinopoulou et al.\cite{andrinopoulou2020integrating} integrated latent classes in the Bayesian shared parameter joint model. The joint model for continuous longitudinal data incorporating random effects into the latent class model in a frequentist framework was proposed by Liu et al.\cite{liu2015joint}. Besides different estimation approaches, Wong et al.\cite{wong2022semiparametric} proposed a JLCM with shared random effects which allows for multivariate longitudinal biomarkers to study the risk of cardiovascular disease through systolic blood pressure and cholesterol levels. \\
The standard JLCM and the JLCM with incorporated random effects assume that the class membership is static. However, an individual’s health status may evolve considerably over time, necessitating start of therapy, different treatment strategy, or changes in monitoring frequency. Allowing class membership to change over time captures changes in disease course more accurately. 
To study impairment levels in everyday tasks in the elderly, Lin et al.\cite{lin2014dynamic} developed a dynamic trajectory class model for intensively measured longitudinal categorical variables where individuals change classes according to a multinomial logistic regression model. The event submodel models the binary outcome of experiencing the event, rather than the time-to-event outcome. Furthermore, Miao and Charalambous\cite{miao2022joint} proposed a joint latent class model with time-varying class membership. However, they assume independence in class membership along the repeated measurement timepoints and assume that the event outcome only depends on the class at the last timepoint. On the other hand, in a shared parameter joint modelling framework, Bartolucci  and Farcomeni \cite{bartolucci2019shared} proposed a shared random effects continuous time hidden Markov model for longitudinal data with informative dropout. In their work, they allow for time-dependent random effects, with transitions following a continuous-time hidden Markov model. \\
In this paper, we develop a dynamic joint latent class model (DJLCM) in which individuals can transition between classes according to a first-order discrete hidden Markov Model. The longitudinal outcome, given the class at each timepoint, is modelled using a class-specific linear mixed effects model. We include a survival outcome, where conditional on the class, the hazard of event follows a Cox proportional hazards model. A shared random effects association structure is used to capture the association between the longitudinal and survival outcomes. Rather than relying on the traditional likelihood function for inference, we use the forward algorithm \cite{baum1970maximization, rabiner1989tutorial}, an iterative algorithm commonly used in hidden Markov Models to efficiently compute the likelihood. We validate our methodology via simulation studies, and apply our model to a clinical trial on patients with metastatic colorectal cancer.\\
This article is structured as follows. Section 2 outlines the novel methodology, starting with the class transition model, followed by the longitudinal and survival models, Section 3 presents the Bayesian implementation approach, specifically the likelihood function via the forward algorithm and prior elicitation for inference. Section 4 discusses methods for posterior classification of subjects.  Section 5 provides simulation studies where we investigate both parameter estimation and classification in our proposed DJLCM. Section 6 presents the data application, including model specification and findings.  Finally, Section 7 concludes with a summary of main contributions, main findings in the data application, limitations of the current work, and directions for future research.

\section{Model Specification}
\subsection{Notation}
Consider a study comprising $N$ independent subjects. Let $T_i^*$ denote the true failure time for subject $i$. The observed event time is defined as $T_i = \min(T_i^*, \mathcal{C}_i)$, 
where $\mathcal{C}_i$ denotes the censoring time. The event indicator is given by
$\Delta_i = \mathbf{1}\{T_i^* \le \mathcal{C}_i\}$,
which equals 1 if the event occurs prior to censoring and 0 otherwise.
\\
For the longitudinal process, let $y_{ij} = y_i(s_{ij})$ denote the $j$th measurement for subject $i$ taken at time $s_{ij}$, corresponding to the $j$th repeated observation, $j=1,2,\cdots,m_i,~i=1,2,\cdots,N$. We assume independence across subjects. The observed data consist of
$\{T_i, \Delta_i, y_{ij} : i = 1, \ldots, N;\; j = 1, \ldots, m_i\}$.
\\
In the following, we introduce our DJLCM, which incorporates transition probabilities. The DJLCM consists of three parts: (1) the time-dependent latent class membership probabilities, (2) the longitudinal submodel, and (3) the survival submodel.
\subsection{Time-Dependent Latent Class Membership Probabilities}
Let $C_i(s) \in \{1, \ldots,  G \}$ denote the latent class membership of subject $i$ at time $s$. We use a multinomial logistic regression for both initial class membership and transitions. At the initial observation time $s_{i1}$, the class membership is modelled as
\begin{equation}\label{cls_init}
P(C_i(s_{i1}) = g) = \frac{\exp(\bm{\delta}_g^{\top} \bm{v}_{i1})}{\sum_{h=1}^{G} \exp(\bm{\delta}_h^{\top} \bm{v}_{i1})}, 
\quad g = 1,\ldots,G,
\end{equation}
where $\bm{v}_{i1}$ are baseline covariates, $\bm{\delta}_{g}$ is the vector of class-specific regression coefficients for initial class membership, and one class (commonly $g=G$) as the reference class with $\bm{\delta}_G = 0$ for identifiability. \\
 For subsequent time points $s_{ij}$, $j = 2, \ldots, m_i$, the latent class process is assumed to follow a first-order discrete-time Markov model:
\begin{equation}\label{cls_trans}
 \pi_{gl}=P(C_i(s_{ij}) = l \mid C_i(s_{i,j-1}) = g) 
= \frac{\exp(\bm{\eta_{gl}}^{\mathsf T} \bm{v}_{ij})}{\sum_{h=1}^{G} \exp(\bm{\eta_{gl}}^{\mathsf T} \bm{v}_{ij})}, 
\quad g,l = 1,\ldots,G,
\end{equation}
 where $\bm{v}_{ij}$ is a vector of possibly time-dependent covariates,  $\bm{\eta}_{gl}$ denotes the class-specific regression coefficient vector governing the probability of transitioning from class $g$ to class $l$, with $\bm{\eta}_{gG} = \mathbf{0}$ for all $g$ to ensure identifiability.
\\
This framework allows latent class memberships to evolve over time under a time-dependent Markovian structure, influenced by both baseline and time-dependent covariates. The latent class process $\{C_i(s_{ij})\}_{j=1}^{m_i}$ can thus be interpreted as a Hidden Markov Model (HMM), where the latent states correspond to class memberships and the observed longitudinal outcomes serve as the emissions \cite{rabiner1989tutorial, jacqmin2010score}. Conditional on the hidden states and class-specific random effects, the observations are independent, and the likelihood can be efficiently computed using standard HMM techniques such as the forward algorithm. 

\subsection{Longitudinal Model}\label{longitudinal}
Conditional on the current latent class $C_i(s_{ij}) = g$, the longitudinal outcome $y_i(s_{ij})$ is modelled using a class-specific linear mixed-effects model:
\begin{eqnarray}\label{long}
y_i(s_{ij}) \,\Big|\, _{C_i(s_{ij}) = g}
= \bm{x}_{i}(s_{ij})^\mathsf{T} \bm{\beta}_g
+ \bm{z}_{i}(s_{ij})^\mathsf{T}\bm{b}_{ig} 
+ \epsilon_{ijg},
\end{eqnarray}
where  $\bm{x}_{i}(s_{ij})$ is the vector of fixed-effect covariates at time $s_{ij}$, with corresponding class-specific coefficients $\bm{\beta}_g$,
    $\bm{z}_{i}(s_{ij})$ is the vector of covariates associated with the random effects $\bm{b}_{ig} \sim \mathcal{N}(0, \bm{D}_g)$, representing subject-specific deviations for class $g$ and
 $\epsilon_{ijg} \sim \mathcal{N}(0, \sigma_g^2)$ is the residual error term.
\\
The corresponding conditional density of $y_i(s_{ij})$ given the latent class and random effects is
\[
f_y\big(y_i(s_{ij}) \mid C_i(s_{ij}) = g,\bm{b}_{ig} \big) = 
\mathcal{N}\Big(\bm{x}_{i}(s_{ij})^\mathsf{T} \bm{\beta}_g+ \bm{z}_{i}(s_{ij})^\mathsf{T}\bm{b}_{ig}, \, \sigma_g^2 \Big).
\]

\subsection{Survival Model}\label{survival}

The latent class process $C_{i}(s)$ is specified as piecewise-constant,
motivated by the discrete nature of longitudinal follow-up: since class
membership can only be inferred at observation times $s_{ij}$, assuming
constancy between visits is both natural and parsimonious
\cite{bartolucci2013}. This construction is consistent with the HMM interpretation introduced in Section~\ref{longitudinal},
where latent states are updated only at observation times
\cite{jackson2011}. Furthermore, such piecewise specifications are
standard practice in joint modelling contexts \cite{rizopoulos2012}.
\\
By construction, the latent class process $C_{i}(s)$ is piecewise-constant,
defined as follows:
\begin{equation*}\label{C_definition}
C_i(s)=
\begin{cases}
C_i(s_{ij}) = C_{ij} & \text{for } s \in [s_{ij},\, s_{i,j+1});
\quad  j=1,2,\ldots, m_i-1, \\[4pt]
C_i(s_{im_i}) = C_{im_i} & \text{for } s \in [s_{im_i},\, T_i].
\end{cases}
\end{equation*}
This means $C_{i}(s)$ only jumps at the longitudinal observation times and the class
membership vector for subject $i$ is
$\bm{C}_i= (C_{i1}, C_{i2}, \ldots, C_{im_i})$.
Conditional on the latent class trajectory $\bm{C}_i$ and random
effects $\bm{b}_i=\{\bm{b}_{ig}\}_{g=1}^G$, the hazard function for the event time $T_i$
is modelled as
\begin{equation}\label{haz_gdjlcm}
\lambda_i(t \mid C_i(t)=g,\bm{b}_{i})
= \lambda_{0g}(t) \exp\big(\bm{w}_i^\mathsf{T} \bm{\gamma}_{g}
+ \bm{\alpha}_{g}^\mathsf{T}\bm{b}_{ig}\big),
\end{equation}
where $\lambda_{0g}(t)$ is the class-specific baseline hazard,
$\bm{w}_i$ is the vector of baseline covariates, $\bm{\gamma}_{g}$
the corresponding regression coefficients, and $\bm{\alpha}_{g}$
the association parameters linking the random effects
$\bm{b}_{ig}$ to the survival outcome.\\
 Due to the piecewise-constant latent class process, the cumulative hazard 
at the observed time $T_i$ can be decomposed over the intervals defined 
by the latent class change points as
\begin{align}\label{cumhaz}
H_i(T_i \mid \bm{C}_i,\bm{b}_i)
&= \int_0^{T_i} \lambda_i(s \mid \bm{C}_i,\bm{b}_i)\,\mathrm{d}s \nonumber\\
&= \sum_{j=1}^{m_i}
\int_{s_{ij}}^{s_{i,j+1}}
\lambda_i(s \mid C_i(s)=C_{ij},\bm{b}_{iC_{ij}})\,\mathrm{d}s,
\end{align}
where $s_{i1}=0$, $s_{i(m_i+1)}=T_i$, and $C_{ij} = g$ denotes the 
class label of subject $i$ at observation time $s_{ij}$.
The corresponding conditional contribution to the likelihood is
\begin{equation*}\label{surv_contrib}
f(T_i, \Delta_i \mid \bm{C}_i, \bm{b}_i)
= \lambda_i(T_i \mid \bm{C}_i, \bm{b}_i)^{\Delta_i}
\exp\big(-H_i(T_i \mid \bm{C}_i, \bm{b}_i)\big).
\end{equation*}

 \section{Bayesian Implementation}
 \subsection{Likelihood function}
The parameter vector $\bm{\Theta}$ is given by 
$\{\bm{\delta}_g\}_{g=1}^G$, the transition parameters 
$\{\bm{\eta}_{gh}\}_{g \neq h}$, and the class-specific parameters 
$\{\beta_g, \bm{D}_g, \sigma_g, \lambda_{0g}, \gamma_g, \alpha_g\}_{g=1}^G$, 
as defined in equations \eqref{cls_init}, \eqref{cls_trans}, 
\eqref{long}, and \eqref{haz_gdjlcm}. Moreover, the random effects $\bm{b}_{ig}$ and the class membership 
$\bm{C}_i = (C_i(s_{i1}), \ldots, C_i(s_{im_i}))$ are latent variables. 
We introduce the simplified notation $y_{ij} = y_i(s_{ij})$.
\\
The joint distribution of the longitudinal outcomes 
$\bm{y}_i = (y_{i1}, \ldots, y_{im_i})$, 
the event time $T_i$, and the event indicator $\Delta_i$, given the random effects 
$\{\bm{b}_{ig}\}_{g=1}^G$, is
\begin{align}\label{like}
f(\bm{y}_i, T_i, \Delta_i \mid \{\bm{b}_{ig}\}_{g=1}^G) 
&= \sum_{C_{im_i}=1}^{G} \cdots \sum_{C_{i1}=1}^{G} 
    f(\bm{y}_i, T_i, \Delta_i, \bm{C}_i \mid \{\bm{b}_{ig}\}_{g=1}^G) \notag\\
&= \sum_{C_{im_i}=1}^{G} \cdots \sum_{C_{i1}=1}^{G} 
    f(\bm{y}_i \mid \bm{C}_i, \{\bm{b}_{ig}\}_{g=1}^G)\,
    f(T_i, \Delta_i \mid \bm{C}_i, \{\bm{b}_{ig}\}_{g=1}^G)\,
    P(\bm{C}_i) \notag\\
&= \sum_{C_{im_i}=1}^{G} \cdots \sum_{C_{i1}=1}^{G} 
   \Bigg( \prod_{j=1}^{m_i} f(y_{ij} \mid C_{ij}, \bm{b}_{iC_{ij}}) \Bigg)\,
    f(T_i, \Delta_i \mid \bm{C}_i, \{\bm{b}_{ig}\}_{g=1}^G)
    \notag\\
&\qquad\qquad\qquad\qquad\qquad\qquad
    \times\, P(C_{i1})\prod_{j=2}^{m_i} \pi_{C_{i,j-1}C_{ij}}, 
\end{align}
where the second equality follows from the conditional independence assumption, i.e., conditional on the class membership $\bm{C}_i$ and the random effects $\{\bm{b}_{ig}\}_{g=1}^G$, the longitudinal outcomes and the event time are independent \cite{andrinopoulou2020integrating}. Moreover, in the final equation, the longitudinal measurements are assumed independent given the class-specific random effects.
\\
The summation in \eqref{like} involves $G^{m_i}$ terms for patient $i$, representing 
all possible latent class sequences across the $m_i$ observation times, 
making direct evaluation computationally infeasible for even moderate 
$G$ and $m_i$. Exploiting the first-order Markov structure of 
$\bm{C}_i$, the summation can be evaluated efficiently using the 
forward algorithm \cite{baum1970maximization, rabiner1989tutorial}, 
a standard recursive technique in HMMs that avoids explicit enumeration 
of all $G^{m_i}$ configurations, reducing the computational complexity 
from $O(G^{m_i})$ to $O(G^2 m_i)$.

\subsection{Forward algorithm for the dynamic JLCM}
The forward algorithm is a recursive method for computing the likelihood 
in HMMs, avoiding explicit enumeration of all possible latent class 
sequences \cite{rabiner1989tutorial}.  For notational 
convenience, we define the following class-specific quantities:
\begin{align*}
f_{ijg} &= P(y_{ij} \mid C_{ij}=g, \bm{b}_i)  \text{ and}\\
S_{ijg} &= P(T_i > s_{i,j+1} \mid T_i > s_{ij}, C_{ij}=g, \bm{b}_i)
\end{align*}
denoting the class-specific longitudinal density and conditional survival 
probability, respectively.
\\
 The forward probability for subject $i$ at observation 
time $j$ in class $g$ is defined as
\begin{equation*}
\alpha_{ij}(g) = P\!\left(
T_i > s_{i,j+1},\, y_{i1},\ldots,y_{ij},\, C_{ij}=g \mid \bm{b}_i
\right),
\end{equation*}
that is, the probability of observing the longitudinal outcomes up to 
time $s_{ij}$, being in class $g$ at that time, and surviving past 
$s_{i,j+1}$.

\paragraph{Initialisation}\hspace{-0.5em}($j=1$).
At the first observation time, using conditional independence:
\begin{equation*}
\alpha_{i1}(g) = f_{i1g}\, S_{i1g}\, P(C_{i1}=g).
\end{equation*}

\paragraph{Recursion}\hspace{-0.5em}($j=2,\ldots,m_i-1$).
At intermediate time points, summing over all possible previous classes:
\begin{equation*}
\alpha_{ij}(g) = f_{ijg}\, S_{ijg} 
\sum_{l=1}^{G} \pi_{lg}\, \alpha_{i,j-1}(l).
\end{equation*}

\paragraph{Termination}\hspace{-0.5em} ($j=m_i$).
At the final observation time, the survival contribution incorporates 
the event indicator $\Delta_i$:
\begin{equation*}
\alpha_{im_i}(g) = f_{im_ig}\, 
\lambda_i(T_i \mid C_{im_i}=g, \bm{b}_i)^{\Delta_i}\,
S_{im_ig}
\sum_{l=1}^{G} \pi_{lg}\, \alpha_{i,m_i-1}(l).
\end{equation*}

\paragraph{Special case}\hspace{-0.5em}($m_i=1$).
For subjects with a single observation:
\begin{equation*}
\alpha_{i1}(g) = f_{i1g}\,
\lambda_i(T_i \mid C_{i1}=g, \bm{b}_i)^{\Delta_i}\,
S_{i1g}\,
P(C_{i1}=g).
\end{equation*}

In compact form:
\begin{equation}\label{fw_algo}
\alpha_{ij}(g)=
\begin{cases}
f_{i1g}\, S_{i1g}\, P(C_{i1}=g),
& j=1, \\[1ex]
\displaystyle f_{ijg}\, S_{ijg}
\sum_{l=1}^{G} \pi_{lg}\, \alpha_{i,j-1}(l),
& j=2,\ldots,m_i-1, \\[1ex]
\displaystyle f_{im_ig}\,
\lambda_i(T_i \mid C_{im_i}=g, \bm{b}_i)^{\Delta_i}\,
S_{im_ig}
\sum_{l=1}^{G} \pi_{lg}\, \alpha_{i,m_i-1}(l),
& j=m_i\neq 1, \\[1ex]
f_{i1g}\,
\lambda_i(T_i \mid C_{i1}=g, \bm{b}_i)^{\Delta_i}\,
S_{i1g}\, P(C_{i1}=g),
& m_i=1,
\end{cases}
\end{equation}
where $s_{i,m_i+1}=T_i$, and the conditional survival probability $S_{ijg}$, in accordance 
with \eqref{cumhaz}, reduces to
\begin{align*}\label{cumhaz_cond}
S_{ijg} 
&= P\big(T_i > s_{i,j+1} \mid T_i > s_{ij},\, C_{ij}=g,\, \bm{b}_i\big) 
\nonumber \\
&= \exp\Bigg(-\int_{s_{ij}}^{s_{i,j+1}} \lambda_{0g}(s)\,
   \exp\!\big(\bm{w}_i^{\mathsf{T}} \bm{\gamma}_{g}
   + \bm{\alpha}_{g}^{\mathsf{T}} \bm{b}_{ig}\big)\,\mathrm{d}s\Bigg).
\end{align*}
Finally, the likelihood for subject $i$ is obtained by summing the 
forward probabilities over all classes at the final observation time:
\begin{equation}\label{lik_forward}
L_i(\bm{\Theta}) = \sum_{g=1}^{G} \alpha_{im_i}(g).
\end{equation}

\subsection{Posterior inference}
We perform Bayesian inference for the dynamic JLCM by combining the 
likelihood $L_i(\bm{\Theta})$ defined in \eqref{lik_forward} with 
prior distributions on all unknown parameters. The joint posterior 
distribution is obtained via Bayes' theorem, and posterior summaries 
are computed using HMC sampling.

\subsubsection{Prior elicitation}
We specify priors for all unknown parameters. Let $g = 1, \ldots, G$ 
and $l = 1, \ldots, G$. The priors are
\begin{align*}\label{priors}
\bm{\delta}_g &\sim \mathcal{N}(\bm{\mu_{\delta g}}, \bm{\Sigma_{\delta g})},\\
\bm{\eta}_{gl} &\sim \mathcal{N}(\bm{\mu_{\eta gl}}, \bm{\Sigma_{\eta gl})},\\
\bm{\beta}_g &\sim \mathcal{N}(\bm{\mu_{\beta g}}, \bm{\Sigma_{\beta g})},\\
\bm{D}_g &= \bm{\Phi}_g \bm{R}_g \bm{\Phi}_g = \bm{\Phi}_g \bm{L}_g \bm{L}_g^\top \bm{\Phi}_g, 
\quad \bm{L}_g \sim \text{LKJ-Cholesky}(\zeta_g),\\
\bm{\Phi}_g &= \mathrm{diag}(\tau_{g1}, \ldots, \tau_{gq}), 
\quad \tau_{gj} \sim \mathrm{half\text{-}Cauchy}(0, \nu_g),\\
\sigma_g &\sim \mathrm{half\text{-}Cauchy}(0, \psi_g),\\
\bm{\alpha}_g &\sim \mathcal{N}(\bm{\mu_{\alpha g}}, 
\bm{\Sigma_{\alpha g})},\\
\bm{\gamma}_g &\sim \mathcal{N}(\bm{\mu_{\gamma g}}, 
\bm{\Sigma_{\gamma g})},\\
\lambda_{0g} &\sim \mathrm{Gamma}(\omega_g, \xi_g).
\end{align*}
where $\text{LKJ-Cholesky}$ denotes the LKJ prior \cite{lewandowski2009generating} on the Cholesky
factor of the correlation matrix of random effects, and $\mathrm{half\text{-}Cauchy}$ denotes
the half-Cauchy prior distribution \cite{gelman2006prior}.
The hyperparameters are chosen to yield weakly uninformative priors, and the model parameters were assumed to be a priori independent.
\\
The joint posterior distribution of $\bm{\Theta}$ and the random 
effects $\{\bm{b}_{ig}\}_{i=1}^N$ given all observed data 
$\{\bm{y}_i, T_i, \Delta_i\}_{i=1}^N$ is
\begin{equation}\label{posterior}
\pi\big(\bm{\Theta}, \{\bm{b}_{ig}\}_{i=1}^N \mid 
\{\bm{y}_i, T_i, \Delta_i\}_{i=1}^N\big)
\propto \prod_{i=1}^N L_i(\bm{\Theta})\, 
\prod_{g=1}^G p(\bm{b}_{ig})\,
\pi(\bm{\Theta}),
\end{equation}
where $\pi(\bm{\Theta}) = \prod_{k} \pi(\theta_k)$ reflects the 
assumption of independent priors for the components $\theta_k$ of 
$\bm{\Theta}$, and $L_i(\bm{\Theta})$ is the likelihood 
for subject $i$ defined in \eqref{lik_forward}.
 We employ HMC using the \texttt{cmdstanr} package 
 \cite{cmdstanr}, which provides efficient and scalable inference 
 for HMC.

\section{Posterior classification}
One important aspect of the DJLCM is posterior classification of 
subjects. Posterior class membership is defined as the probability 
of a subject belonging to a particular class given all observed 
longitudinal and survival data \cite{proust2014joint}. Due to time-dependent class membership 
in our DJLCM, posterior class probabilities are computed at each 
timepoint, and subjects are assigned to the class with the maximum 
posterior probability. \\
Let $ \bm{\mathcal{Y}}(s_{ij})=\{ y_i(s_{il}); l=1,\ldots , j\}$ denote all the longitudinal measurements for subject $i$ up to timepoint $j$.
We begin by deriving the posterior class probability conditionally on 
the random effects $\bm{b}_i$. Using Bayes' theorem and marginalising 
over all possible past class sequences, the conditional posterior class 
probability at observation time $j$ for subject $i$ is:
\begin{align}\label{post_class_cond}
&P(C_{ij}=g \mid \bm{\mathcal{Y}}(s_{ij}), T_i, \Delta_i, \widehat{\bm{b}}_i;\, 
\widehat{\bm{\Theta}}) \nonumber\\[6pt]
&= \frac{P(\bm{\mathcal{Y}}(s_{ij}), T_i, \Delta_i, C_{ij}=g \mid \widehat{\bm{b}}_i;\, 
\widehat{\bm{\Theta}})}{P(\bm{\mathcal{Y}}(s_{ij}), T_i, \Delta_i \mid \widehat{\bm{b}}_i;\, 
\widehat{\bm{\Theta}})} \nonumber\\[6pt]
&= \frac{
\displaystyle\sum_{C_{i(j-1)}=1}^{G}\cdots\sum_{C_{i1}=1}^{G}
P(\bm{\mathcal{Y}}(s_{ij}), T_i, \Delta_i, C_{ij}=g, C_{i,j-1}, \ldots, C_{i1} 
\mid \widehat{\bm{b}}_i;\, \widehat{\bm{\Theta}})
}{
\displaystyle\sum_{C_{ij}=1}^{G}\sum_{C_{i,j-1}=1}^{G}
\cdots\sum_{C_{i1}=1}^{G}
P(\bm{\mathcal{Y}}(s_{ij}), T_i, \Delta_i, C_{ij}, C_{i,j-1}, \ldots, C_{i1} 
\mid \widehat{\bm{b}}_i;\, \widehat{\bm{\Theta}})
} \nonumber\\[6pt]
&= \frac{\alpha_{ij}(g \mid \widehat{\bm{b}}_i)}
{\sum_{l=1}^{G}\alpha_{ij}(l \mid \widehat{\bm{b}}_i)},
\end{align}
for $j=1,2,\ldots,m_i$, where $\alpha_{ij}(g \mid \widehat{\bm{b}}_i)$ is the 
forward probability defined in \eqref{fw_algo}. The final equality 
follows directly from the recursive structure of the forward algorithm.
\\
Since $\bm{b}_i$ is a latent quantity, the fully Bayesian posterior 
class probability is obtained by integrating \eqref{post_class_cond} 
over the posterior distribution of $\bm{b}_i$:
\begin{align*}\label{post_class}
\pi_{ijg}^{Y,T}
&= P(C_{ij}=g \mid y_{i\{1:j\}}, T_i, \Delta_i;\, 
\widehat{\bm{\Theta}}) \nonumber\\[6pt]
&= \int P(C_{ij}=g \mid y_{i\{1:j\}}, T_i, \Delta_i, \bm{b}_i;\, 
\widehat{\bm{\Theta}})\,
p(\bm{b}_i \mid \bm{\mathcal{Y}}(s_{ij}), T_i, \Delta_i;\, 
\widehat{\bm{\Theta}})\, \mathrm{d}\bm{b}_i \nonumber\\[6pt]
&= \int 
\frac{\alpha_{ij}(g \mid \bm{b}_i)}
{\sum_{l=1}^{G}\alpha_{ij}(l \mid \bm{b}_i)}\,
p(\bm{b}_i \mid \bm{\mathcal{Y}}(s_{ij}), T_i, \Delta_i;\, 
\widehat{\bm{\Theta}})\, \mathrm{d}\bm{b}_i \nonumber\\[6pt]
&\approx \frac{1}{M}\sum_{m=1}^{M} 
\frac{\alpha_{ij}(g \mid \bm{b}_i^{(m)})}
{\sum_{l=1}^{G}\alpha_{ij}(l \mid \bm{b}_i^{(m)})},
\end{align*}
for $j=1,2,\ldots,m_i$, where $\bm{b}_i^{(m)}$ denotes the $m$-th 
posterior MCMC draw of the random effects for subject $i$ obtained 
from \eqref{posterior}. The approximation follows from replacing the 
intractable integral with a Monte Carlo average over $M$ posterior 
draws.
\\
Finally, subject $i$ is assigned to class $\hat{C}_{ij}$ at 
observation time $j$ according to the maximum a posteriori (MAP) rule:
\begin{equation*}\label{map_class}
\hat{C}_{ij} = \operatorname*{arg\,max}_{g \in \{1,\ldots,G\}} \pi_{ijg}^{Y,T},
\quad j=1,2,\ldots,m_i.
\end{equation*}

\section{Simulation Studies}
In this section, we evaluate the performance of the proposed model through simulation studies involving two-class and three-class dynamic joint latent class models. In these models, class transitions for each subject over successive observation times are governed by a HMM with specified transition probabilities. We investigate the model's performance in terms of parameter estimation and latent class classification. Parameter estimation is assessed using bias, precision, and coverage, while classification performance is evaluated using measures of classification accuracy and the certainty of class assignment.

\subsection{Scenario 1: Two latent classes}

\subsubsection{Simulation design}

We consider $N=1000$ individuals and perform $100$ simulation replications. For each replication, we generate longitudinal and survival data from a DJLCM with $G=2$ latent classes. The latent class membership evolves over time according to a first-order HMM, with transition probabilities depending on a binary covariate.

\subsubsection*{Initial latent class membership and transition model}

At baseline, the latent class membership $C_i(0)$ is generated according to a Bernoulli distribution:
\[
  C_i(0) \sim \operatorname{Bernoulli}(p_i),
\]
where $C_i(0)=1$ indicates membership in Class~1, and Class~2 is the reference class. The probability of belonging to Class~1 is specified through a logistic regression model:
\[
  \operatorname{logit}(p_i) = \delta_{10} + \delta_{11} x_{1i},
\]
where $x_{1i} \sim \operatorname{Bernoulli}(0.5)$, with $\delta_{10}=-2$ and $\delta_{11}=4$.
\\
For subsequent observation times $s_{ij}$, the latent class membership evolves according to a first-order HMM. Specifically, the transition probability from the previous class $g$ to Class~1 is given by
\begin{align*}
  \pi_{1g} = P(C_i(s_{ij}) = 1 \mid C_i(s_{i,j-1}) = g) 
= \frac{\exp(\eta_{g0} + \eta_{g1} \, z_{i})}{1 + \exp(\eta_{g0} + \eta_{g1} \, z_{i})},
\end{align*}
where $z_i \sim \operatorname{Bernoulli}(0.5)$. The probability of transitioning to Class~2 is $\pi_{g2}(s_{ij}) = 1 - \pi_{g1}(s_{ij})$. The transition parameters are set to $(\eta_{10},\eta_{11})=(0.4,1)$ for subjects previously in Class~1 and $(\eta_{20},\eta_{21})=(-0.2,2)$ for subjects previously in Class~2.

\subsubsection*{Longitudinal submodel}

The longitudinal outcome $y_{ij}$ for subject $i$, observed at time $s_{ij}$ and belonging to latent class $g$, is generated from a linear mixed-effects model:
\begin{equation*}
y_{ij}\mid _{ C_{ij}=g} = \beta_{g0} + \beta_{g1} \, s_{ij} + \beta_{g2} \, x_{1i} + \beta_{g3} \, x_{2i} + b_{ig0} + b_{ig1}s_{ij} + \epsilon_{ijg},
\end{equation*}
where $x_{2i} \sim \mathcal{N}(0,1)$, and the measurement errors satisfy $\epsilon_{ijg} \sim \mathcal{N}(0,\sigma_g^2)$. The class-specific random intercepts and slopes are represented by $\mathbf{b}_{ig} = (b_{ig0}, b_{ig1})^\top$, with
\[
  \mathbf{b}_{ig} \sim \mathcal{N}(\mathbf{0}, \mathbf{D}_g), \qquad
  \mathbf{D}_g =
  \begin{pmatrix} 1 & 0.5 \\ 0.5 & 1 \end{pmatrix}.
\]
The true regression coefficients are set to $\beta_{gk}=-2$ for $g=1$ and $\beta_{gk}=2$ for $g=2$, for $k=0,1,2,3$. The residual standard deviations are $\sigma_1=1$ and $\sigma_2=2$.
\\
Subjects are followed for up to two years, with unbalanced longitudinal measurements and a maximum of $21$ observations per subject.

\subsubsection*{Survival submodel}

The event time $T_i$ is generated from a class-specific proportional hazards model that incorporates baseline covariates and the shared random effects from the longitudinal submodel. The hazard function is specified as
\begin{equation*}
  \lambda_i(t \mid C_i(t)=g, \mathbf{b}_i)
  = \lambda_{0g}(t)
    \exp\Big\{
      \gamma_{g0} + \gamma_{g1} w_{1i} + \gamma_{g2} w_{2i}
      + \alpha_{g0} b_{ig0} + \alpha_{g1} b_{ig1}
    \Big\},
\end{equation*}
where $w_{1i}, w_{2i} \sim \operatorname{Bernoulli}(0.5)$, and $\lambda_{0g}(t)$ denotes the class-specific Weibull baseline hazard, with shape parameters $\omega_1=1.5$ and $\omega_2=2$.
\\
The survival regression coefficients are set to $(\gamma_{10},\gamma_{11},\gamma_{12})=(-0.7,-1.2,-0.5)$ for Class~1 and $(\gamma_{20},\gamma_{21},\gamma_{22})=(-0.3,-0.8,-0.5)$ for Class~2. The associations between the longitudinal random effects and the survival process are specified as $(\alpha_{10},\alpha_{11})=(0.5,0.5)$ for Class~1 and $(\alpha_{20},\alpha_{21})=(0.1,0.1)$ for Class~2.
\\
Independent censoring times are generated from an exponential distribution with rate $0.1$. Administrative censoring is imposed at two years. The observed follow-up time and event indicator are therefore
$  \widetilde{T}_i = \min(T_i,\mathcal{C}_i, 2),
~
  \Delta_i = \mathbf{1}\{T_i \leq \min(\mathcal{C}_i, 2)\},$
where $\Delta_i=1$ indicates an observed event and $\Delta_i=0$ indicates censoring.
\\
Figure~\ref{individual_trajectories_K2} displays the longitudinal trajectories of five randomly selected subjects generated under the DJLCM with $G=2$. Blue and orange points represent observations belonging to Classes~1, and 2, respectively. Observed event times are indicated by red lines, with solid lines denoting events and dashed lines denoting censoring.

\subsubsection*{Prior elicitation}
We used uninformative priors for all model parameters, as described below.

\begin{equation*}
\begin{alignedat}{2}
\delta_{g0},\ \delta_{g1} &\sim \mathcal{N}(0, 100), &\qquad & g \neq 2,\\
\eta_{gl0},\ \eta_{gl1} &\sim \mathcal{N}(0, 100), && l \neq 2,\\
\beta_{g0},\ \beta_{g1},\ \beta_{g3},\ \beta_{g4} &\sim \mathcal{N}(0, 100), &&\\
\gamma_{g0},\ \gamma_{g1},\ \gamma_{g2} &\sim \mathcal{N}(0, 100), &&\\
\alpha_{g0},\ \alpha_{g1} &\sim \mathcal{N}(0, 100), &&\\
\bm{L}_g &\sim \text{LKJ-Cholesky}(2), &&\\
\tau_{gj} &\sim \mathrm{half\text{-}Cauchy}(0, 5), && j = 1, 2,\\
\sigma_g &\sim \mathrm{half\text{-}Cauchy}(0, 5), &&\\
\omega_g &\sim \mathrm{Gamma}(0.01, 0.01), &&
\end{alignedat}
\end{equation*}
where $g = 1, 2$ unless stated otherwise. \\
For implementation, we ran $30{,}000$ HMC iterations, including $15{,}000$ iterations as burn-in.  \\

\subsubsection{Performance evaluation}

\subsubsection*{Estimation}

We evaluate the performance of parameter estimation across the $R=100$ simulation replications using the average estimate, standard deviation (SD), relative bias (RB), root mean squared error (RMSE), and coverage ratio (CR). Since class labels are arbitrary and may differ across replications, we manually relabel the classes. In each replication we assign each estimated class to the true class whose intercept of the longitudinal submodel is closest to its estimated intercept before computing these summaries.
The results are reported in Table~\ref{simK2_djlcm}. Overall, the estimated values are close to their corresponding true values, with generally small relative biases and RMSEs. The coverage ratios are close to the nominal level of $95\%$ for most parameters, indicating satisfactory estimation performance. However, some parameters, particularly $\alpha_{20}$, $\alpha_{21}$, and $\gamma_{22}$, exhibit greater uncertainty or lower coverage, suggesting that these parameters are more difficult to estimate accurately under the considered simulation setting. Nevertheless, the results generally demonstrate that the proposed DJLCM provides reliable parameter estimates.

\subsubsection*{Classification}
We assess latent class classification by comparing the estimated class assignments with the true latent class memberships at each observation time. 
\\
For the two-class scenario, Class~2 is designated as the positive class and Class~1 as the negative class. Comparing the estimated and true class assignments yields the numbers of true positives (TP), false positives (FP), true negatives (TN), and false negatives (FN). The classification performance measures are defined as follows:
\begin{equation*}
  \mathrm{Sensitivity} = \frac{\mathrm{TP}}{\mathrm{TP}+\mathrm{FN}},~
  \mathrm{Specificity} = \frac{\mathrm{TN}}{\mathrm{TN}+\mathrm{FP}},~
  \mathrm{PPV} = \frac{\mathrm{TP}}{\mathrm{TP}+\mathrm{FP}},~
  \mathrm{NPV} = \frac{\mathrm{TN}}{\mathrm{TN}+\mathrm{FN}}.
\end{equation*}
Sensitivity measures the proportion of observations truly belonging to Class~2 that are correctly classified, whereas specificity measures the proportion of observations truly belonging to Class~1 that are correctly classified. PPV and NPV quantify the proportions of observations assigned to Classes~2 and~1, respectively, that are correctly classified.
\\
Table~\ref{sens_spec_K2} reports the mean and standard deviation of the classification measures across the $100$ simulation replications, while Table~\ref{conf_matrix_K2} presents the confusion matrix. The results indicate good classification performance for both latent classes. The sensitivity is $98.44\%$ for Class~1 and $92.89\%$ for Class~2, indicating that the model identifies observations belonging to Class~1 more accurately than those belonging to Class~2. The specificity values are also high, at $92.89\%$ and $99.87\%$ for Classes~1 and~2, respectively. Both PPV and NPV exceed $96\%$, demonstrating strong agreement between the estimated and true class assignments. The confusion matrix similarly shows that most observations are correctly classified, with $98.44\%$ of true Class~1 observations and $92.90\%$ of true Class~2 observations assigned to their correct classes. Overall, these results indicate that the proposed DJLCM achieves good latent class discrimination under the two-class simulation scenario.

\begin{table}[!htbp]
\centering
\footnotesize
\setlength{\tabcolsep}{3pt}
\caption{
Parameter estimates (Est.), standard deviation (SD), Relative Bias (RB), Root Mean Squared Error (RMSE), and Coverage Ratio (CR) for the simulation study with $G=2$.}
\label{simK2_djlcm}
\begin{adjustbox}{max width=\textwidth}
\begin{tabular}{l rrrrrr @{\hspace{12pt}} rrrrrr}
\toprule
 & \multicolumn{6}{c}{Class 1} & \multicolumn{6}{c}{Class 2} \\
\cmidrule(lr){2-7} \cmidrule(lr){8-13}
 & Real & Est. & SD & RB & RMSE & CR
 & Real & Est. & SD & RB & RMSE & CR \\
\midrule
$\delta_{g0}$ & -2.00 & -2.04 & 0.20 & 0.02 & 0.19 & 0.96
 & \multicolumn{6}{c}{Reference class} \\
$\delta_{g1}$ & 4.00 & 4.04 & 0.25 & 0.01 & 0.24 & 0.98
 & \multicolumn{6}{c}{} \\
\addlinespace
$\eta_{g10}$ & 0.40 & 0.40 & 0.04 & 0.01 & 0.04 & 0.96
 & -0.20 & -0.19 & 0.05 & -0.03 & 0.05 & 0.95 \\
$\eta_{g11}$ & 1.00 & 1.00 & 0.06 & 0.00 & 0.05 & 0.99
 & 2.00 & 2.00 & 0.11 & -0.00 & 0.10 & 0.95 \\
\addlinespace
$\beta_{g0}$ & -2.00 & -2.00 & 0.06 & 0.00 & 0.06 & 0.98
 & 2.00 & 1.99 & 0.09 & -0.00 & 0.09 & 0.90 \\
$\beta_{g1}$ & -2.00 & -2.00 & 0.05 & -0.00 & 0.06 & 0.94
 & 2.00 & 2.00 & 0.09 & 0.00 & 0.09 & 0.95 \\
$\beta_{g2}$ & -2.00 & -1.99 & 0.08 & -0.00 & 0.08 & 0.94
 & 2.00 & 2.00 & 0.12 & 0.00 & 0.12 & 0.95 \\
$\beta_{g3}$ & -2.00 & -2.00 & 0.04 & -0.00 & 0.04 & 0.94
 & 2.00 & 2.00 & 0.06 & -0.00 & 0.06 & 0.95 \\
\addlinespace
$\gamma_{g0}$& -0.70 & -0.74 & 0.11 & 0.06 & 0.12 & 0.94
 & -0.30 & -0.35 & 0.13 & 0.16 & 0.13 & 0.96 \\
$\gamma_{g1}$& -1.20 & -1.21 & 0.14 & 0.01 & 0.13 & 0.97
 & -0.80 & -0.87 & 0.15 & 0.09 & 0.16 & 0.95 \\
$\gamma_{g2}$& -0.50 & -0.49 & 0.13 & -0.02 & 0.12 & 0.94
 & -0.50 & -0.53 & 0.14 & 0.06 & 0.18 & 0.87 \\
\addlinespace
$\alpha_{g0}$ & 0.50 & 0.51 & 0.14 & 0.01 & 0.13 & 0.95
 & 0.10 & 0.11 & 0.68 & 0.14 & 0.46 & 0.95 \\
$\alpha_{g1}$ & 0.50 & 0.53 & 0.16 & 0.05 & 0.16 & 0.96
 & 0.10 & 0.12 & 0.67 & 0.17 & 0.46 & 0.95 \\
\addlinespace
$\sigma_g$ & 1.00 & 1.00 & 0.01 & 0.00 & 0.01 & 0.90
 & 2.00 & 2.00 & 0.03 & 0.00 & 0.03 & 0.94 \\
$\omega_g$ & 1.50 & 1.53 & 0.09 & 0.02 & 0.09 & 0.95
 & 2.00 & 2.07 & 0.13 & 0.04 & 0.14 & 0.94 \\
\addlinespace
$D_{g,11}$ & 1.00 & 1.00 & 0.07 & -0.00 & 0.07 & 0.94
 & 1.00 & 0.97 & 0.18 & -0.03 & 0.18 & 0.90 \\
$D_{g,12}$ & 0.50 & 0.49 & 0.06 & -0.03 & 0.06 & 0.90
 & 0.50 & 0.51 & 0.14 & 0.02 & 0.11 & 0.99 \\
$D_{g,22}$ & 1.00 & 1.02 & 0.09 & 0.02 & 0.10 & 0.90
 & 1.00 & 1.01 & 0.21 & 0.01 & 0.19 & 0.97 \\
\bottomrule
\end{tabular}
\end{adjustbox}
\end{table}

\begin{table}[H]
  \centering
  \caption{Mean (in percentage) and standard deviation (in brackets) of sensitivity, specificity, PPV, and NPV for the two classes for the simulation study with $G=2$.}
  \label{sens_spec_K2}
  \begin{tabular}{c c c c c}
    \toprule
    Class & Sensitivity & Specificity & PPV & NPV \\
    \midrule
    \multirow{2}{*}{1} & 98.44\%     & 92.89\%  & 96.40\%  & 96.85\%  \\
                       & (0.22\%)  & (0.68\%)  & (0.32\%)    & (0.42\%)  \\
    \bottomrule
  \end{tabular}
\end{table}
 
\begin{table}[ht]
\centering
\caption{Confusion matrix: mean (SD) classification rates across 100 simulation replications.}
\label{conf_matrix_K2}
\begin{tabular}{lcc}
\toprule
\multirow{2}{*}{Assigned class} & \multicolumn{2}{c}{True class} \\
\cmidrule(lr){2-3}
 & Class 1 & Class 2 \\
\midrule
Class 1 & $98.44\%\ (0.22\%)$ & $1.56\%\ (0.22\%)$ \\
Class 2 & $7.10\%\ (0.68\%)$  & $92.90\%\ (0.68\%)$ \\
\bottomrule
\end{tabular}
\end{table}

 \subsection{Scenario 2: Three latent classes }

 \subsubsection{Simulation design}
 
The simulation setup for $G=3$ is analogous to that for $G=2$. We consider $N=2000$ individuals and perform $R=100$ simulation replications.

\subsubsection*{Initial latent class membership and transition model}

Initially, each subject $i$ belongs to one of the three latent classes, $g\in\{1,2,3\}$, according to a multinomial logistic regression model. Class~3 is the reference class, and the initial class probabilities are specified as
\begin{equation*}
\log\left(
\frac{P(C_i(s_{i1})=g\mid z_i)}
{P(C_i(s_{i1})=3\mid z_i)}
\right)
=\delta_{g0}+\delta_{g1}z_i,
\qquad g=1,2,
\end{equation*}
where $z_i\sim\operatorname{Bernoulli}(0.5)$ is a binary covariate. The true parameter values are $\delta_{10}=-0.861$, $\delta_{11}=-2$, $\delta_{20}=-0.861$, and $\delta_{21}=2$.
\\
For subsequent observation times $s_{ij}$, $j=2,\ldots,m_i$, latent class membership evolves according to a first-order hidden Markov model (HMM). The transition probabilities are denoted by
\[
\pi_{gl}(s_{ij})
=P\{C_i(s_{ij})=l\mid C_i(s_{i,j-1})=g\},
\qquad g,l\in\{1,2,3\}.
\]
For each previous class $g$, the transition probabilities to Classes~1 and~2 are specified using multinomial logistic models:
\begin{align*}
\pi_{g1}(s_{ij})
&=\frac{\exp(\eta_{g10}+\eta_{g11}z_i)}
{1+\exp(\eta_{g10}+\eta_{g11}z_i)
+\exp(\eta_{g20}+\eta_{g21}z_i)},\\
\pi_{g2}(s_{ij})
&=\frac{\exp(\eta_{g20}+\eta_{g21}z_i)}
{1+\exp(\eta_{g10}+\eta_{g11}z_i)
+\exp(\eta_{g20}+\eta_{g21}z_i)},\\
\pi_{g3}(s_{ij})&=1-\pi_{g1}(s_{ij})-\pi_{g2}(s_{ij}).
\end{align*}
The transition parameters are set to $(\eta_{110},\eta_{111})=(0.4,1)$ and $(\eta_{120},\eta_{121})=(-0.3,2.5)$ for transitions from Class~1; $(\eta_{210},\eta_{211})=(-0.2,2)$ and $(\eta_{220},\eta_{221})=(0.7,3.2)$ for transitions from Class~2; and $(\eta_{310},\eta_{311})=(0.1,1.5)$ and $(\eta_{320},\eta_{321})=(-0.5,0.8)$ for transitions from Class~3.

\subsubsection*{Longitudinal submodel}

For subject $i$ belonging to latent class $g$ at time $s_{ij}$, the longitudinal outcome is generated from a linear mixed-effects model:
\begin{equation*}
y_{ij}\mid _{ C_{ij}=g} = \beta_{g0} + \beta_{g1} \, s_{ij} + \beta_{g2} \, x_{1i} + \beta_{g3} \, x_{2i} + b_{ig0} + b_{ig1}s_{ij} + \epsilon_{ijg},
\end{equation*}
where $x_{1i}\sim\mathcal{N}(0,1)$ and $x_{2i}\sim\mathcal{N}(0,1)$ are covariates, and $\epsilon_{ijg}\sim\mathcal{N}(0,\sigma_g^2)$ is the measurement error. The class-specific random-effects vector is
\[
\mathbf{b}_{ig}=(b_{ig0},b_{ig1})^\top,
\qquad
\mathbf{b}_{ig}\sim\mathcal{N}(\mathbf{0},\mathbf{D}_g),
\]
where
\[
\mathbf{D}_g=
\begin{pmatrix}
1&0.5\\
0.5&1
\end{pmatrix}.
\]
The true regression coefficients are $(\beta_{10},\beta_{11},\beta_{12},\beta_{13})=(-14,-2,-2,-2)$ for Class~1, $(\beta_{20},\beta_{21},\beta_{22},\beta_{23})=(0.1,0.1,0.1,0.1)$ for Class~2, and $(\beta_{30},\beta_{31},\beta_{32},\beta_{33})=(14,3,3,3)$ for Class~3. The residual standard deviations are $\sigma_1=1$, $\sigma_2=2$, and $\sigma_3=1$.
\\
Subjects are followed for up to two years, with unbalanced longitudinal measurements and a maximum of 21 observations per subject. Across the 100 datasets, the mean (SD) number of longitudinal measurements per subject is 8.8 (5.5), and the mean (SD) total number of repeated measurements per dataset is 17,648.04 (234.86).

\subsubsection*{Survival submodel}

The event time $T_i$ is generated from a class-specific proportional hazards model incorporating baseline covariates and the shared random effects from the longitudinal submodel. The hazard function is
\begin{equation*}
\lambda_i(t \mid C_i(t)=g ,\bm{b}_{i})
= \lambda_{0g}(t) \exp\big(\gamma_{g0}+\gamma_{g1}w_{1i}+\gamma_{g2}w_{2i}
+ \alpha_{g0}b_{ig0}+\alpha_{g1}b_{ig1}\big),
\end{equation*}
where $\lambda_{0g}(t)$ is a class-specific Weibull baseline hazard with shape parameter $\omega_g$, and $w_{1i},w_{2i}\sim\operatorname{Bernoulli}(0.5)$ are baseline covariates.
\\
For Class~1, the true parameters are $\omega_1=1.5$, $(\gamma_{10},\gamma_{11},\gamma_{12})=(-0.7,-1.2,-0.5)$, and $(\alpha_{10},\alpha_{11})=(0.5,0.5)$. For Class~2, they are $\omega_2=1$, $(\gamma_{20},\gamma_{21},\gamma_{22})=(-0.1,-1,-0.9)$, and $(\alpha_{20},\alpha_{21})=(-0.2,-0.2)$. For Class~3, they are $\omega_3=2$, $(\gamma_{30},\gamma_{31},\gamma_{32})=(-0.3,-0.8,-0.5)$, and $(\alpha_{30},\alpha_{31})=(0.1,0.1)$.
\\
Independent censoring times are generated from an exponential distribution with rate $0.1$. Administrative censoring is imposed at two years. The observed follow-up time and event indicator are defined as in Scenario~1, with an average censoring rate of approximately 50\%.
\\
Figure~\ref{individual_trajectories_K3} displays the longitudinal trajectories of five randomly selected subjects generated under the DJLCM with $G=3$. Blue, orange, and green points represent observations belonging to Classes~1, 2, and 3, respectively. Observed event times are indicated by red lines, with solid lines denoting events and dashed lines denoting censoring.

\subsubsection*{Prior elicitation}
Uninformative priors were used for all model parameters. Specifically,

\begin{equation*}\label{priors_sim}
\begin{alignedat}{2}
\delta_{g0},\ \delta_{g1} &\sim \mathcal{N}(0, 100), &\qquad & g \neq 3,\\
\eta_{gl0},\ \eta_{gl1} &\sim \mathcal{N}(0, 100), && l \neq 3,\\
\beta_{g0},\ \beta_{g1},\ \beta_{g3},\ \beta_{g4} &\sim \mathcal{N}(0, 100), &&\\
\gamma_{g0},\ \gamma_{g1},\ \gamma_{g2} &\sim \mathcal{N}(0, 100), &&\\
\alpha_{g0},\ \alpha_{g1} &\sim \mathcal{N}(0, 100), &&\\
\bm{L}_g &\sim \text{LKJ-Cholesky}(2), &&\\
\tau_{gj} &\sim \mathrm{half\text{-}Cauchy}(0, 5), && j = 1, 2, 3,\\
\sigma_g &\sim \mathrm{half\text{-}Cauchy}(0, 5), &&\\
\omega_g &\sim \mathrm{Gamma}(0.01, 0.01), &&
\end{alignedat}
\end{equation*}
where $g = 1, 2, 3$ unless stated otherwise. 
We used $30{,}000$ HMC iterations, including $15{,}000$ iterations as warm-up.  \\

\subsubsection{Performance evaluation}

\subsubsection*{Estimation}
The results are reported in Table~\ref{simK3_djlcm}.
Overall, the estimated values are close to their corresponding true values, with generally small  RBs and  RMSEs. The coverage ratios  are close to the nominal 95\% level for most parameters, although some departures are observed, particularly for $\eta_{g20}$ in Class~1, $\eta_{g21}$ in Class~3, $\beta_{30}$, $\gamma_{10}$, $\gamma_{20}$, and $\omega_2$. Nevertheless, the overall results indicate satisfactory parameter estimation performance under the considered simulation setting.

\subsubsection*{Classification}
Latent class classification is evaluated by comparing the estimated class assignments with the true latent class memberships at each observation time. For the three-class setting, sensitivity, specificity, PPV, and NPV  are calculated separately for each class using a one-versus-rest approach. Specifically, for each class, sensitivity measures the proportion of observations truly belonging to that class that are correctly assigned to it, while specificity measures the proportion of observations belonging to either of the other two classes that are correctly classified as not belonging to the class of interest. PPV and NPV measure the corresponding proportions among observations assigned to the class of interest and those assigned to the other classes, respectively.
\\
Table~\ref{sens_spec} reports the mean and standard deviation of these measures across the 100 simulation replications. All four measures exceed 99.7\% for each class, indicating excellent classification performance. Table~\ref{conf_matrix} provides a complementary summary of the agreement between estimated and true class memberships. The percentages in the confusion matrix should be interpreted according to their specified denominator and verified against the replication-level results. Overall, the reported classification metrics indicate excellent discrimination among the three latent classes under the considered simulation setting.

\begin{sidewaystable}
\centering
\footnotesize
\setlength{\tabcolsep}{3pt}
\caption{Parameter estimates (Est.), standard deviation (SD), relative bias (RB), root mean squared error (RMSE), and coverage ratio (CR) for the simulation study with $G=3$.}
\label{simK3_djlcm}
\begin{adjustbox}{max width=\linewidth}
\begin{tabular}{l rrrrrr @{\hspace{12pt}} rrrrrr @{\hspace{12pt}} rrrrrr}
\toprule
& \multicolumn{6}{c}{Class 1}
& \multicolumn{6}{c}{Class 2}
& \multicolumn{6}{c}{Class 3} \\
\cmidrule(lr){2-7} \cmidrule(lr){8-13} \cmidrule(lr){14-19}
& Real & Est. & SD & RB & RMSE & CR
& Real & Est. & SD & RB & RMSE & CR
& Real & Est. & SD & RB & RMSE & CR \\
\midrule
$\delta_{g0}$ & -0.86 & -0.87 & 0.08 & 0.01 & 0.09 & 0.95
& -0.86 & -0.87 & 0.08 & 0.01 & 0.08 & 0.96
& \multicolumn{6}{c}{Reference class} \\
$\delta_{g1}$ & -2.00 & -2.02 & 0.12 & 0.01 & 0.12 & 0.94
& 2.00 & 2.00 & 0.12 & 0.00 & 0.11 & 0.94
& \multicolumn{6}{c}{} \\
\addlinespace
$\eta_{g10}$ & 0.40 & 0.40 & 0.04 & 0.01 & 0.05 & 0.95
& -0.20 & -0.20 & 0.06 & 0.01 & 0.06 & 0.94
& 0.10 & 0.10 & 0.05 & 0.02 & 0.05 & 0.96 \\
$\eta_{g11}$ & 1.00 & 0.99 & 0.11 & -0.01 & 0.12 & 0.96
& 2.00 & 2.01 & 0.12 & 0.01 & 0.12 & 0.97
& 1.50 & 1.53 & 0.13 & 0.02 & 0.15 & 0.94 \\
$\eta_{g20}$ & -0.30 & -0.29 & 0.05 & -0.02 & 0.06 & 0.91
& 0.70 & 0.69 & 0.05 & -0.01 & 0.05 & 0.95
& -0.50 & -0.50 & 0.06 & 0.01 & 0.06 & 0.93 \\
$\eta_{g21}$ & 2.50 & 2.50 & 0.11 & -0.00 & 0.11 & 0.96
& 3.20 & 3.22 & 0.11 & 0.01 & 0.12 & 0.94
& 0.80 & 0.82 & 0.16 & 0.03 & 0.18 & 0.91 \\
\addlinespace
$\beta_{g0}$ & -14.00 & -14.00 & 0.04 & -0.00 & 0.04 & 0.96
& 0.10 & 0.10 & 0.05 & 0.04 & 0.04 & 0.97
& 14.00 & 14.01 & 0.04 & 0.00 & 0.05 & 0.89 \\
$\beta_{g1}$ & -2.00 & -2.01 & 0.05 & 0.00 & 0.05 & 0.94
& 0.10 & 0.09 & 0.06 & -0.07 & 0.05 & 0.99
& 3.00 & 3.01 & 0.06 & 0.00 & 0.06 & 0.94 \\
$\beta_{g2}$ & -2.00 & -2.00 & 0.03 & 0.00 & 0.03 & 0.96
& 0.10 & 0.10 & 0.04 & -0.03 & 0.04 & 0.95
& 3.00 & 3.00 & 0.04 & -0.00 & 0.04 & 0.98 \\
$\beta_{g3}$ & -2.00 & -2.00 & 0.03 & 0.00 & 0.03 & 0.96
& 0.10 & 0.10 & 0.04 & 0.03 & 0.04 & 0.97
& 3.00 & 3.00 & 0.04 & -0.00 & 0.04 & 0.95 \\
\addlinespace
$\gamma_{g0}$ & -0.70 & -0.75 & 0.12 & 0.08 & 0.13 & 0.90
& -0.10 & -0.12 & 0.07 & 0.24 & 0.08 & 0.90
& -0.30 & -0.34 & 0.11 & 0.12 & 0.12 & 0.95 \\
$\gamma_{g1}$ & -1.20 & -1.23 & 0.14 & 0.02 & 0.16 & 0.95
& -1.00 & -1.00 & 0.10 & -0.00 & 0.10 & 0.94
& -0.80 & -0.82 & 0.13 & 0.03 & 0.13 & 0.95 \\
$\gamma_{g2}$ & -0.50 & -0.50 & 0.13 & 0.01 & 0.13 & 0.93
& -0.90 & -0.92 & 0.10 & 0.03 & 0.08 & 0.99
& -0.50 & -0.48 & 0.13 & -0.04 & 0.12 & 0.96 \\
\addlinespace
$\alpha_{g0}$ & 0.50 & 0.54 & 0.20 & 0.08 & 0.20 & 0.94
& -0.20 & -0.21 & 0.32 & 0.05 & 0.20 & 0.98
& 0.10 & 0.10 & 0.22 & -0.01 & 0.20 & 0.96 \\
$\alpha_{g1}$ & 0.50 & 0.49 & 0.23 & -0.02 & 0.25 & 0.93
& -0.20 & -0.20 & 0.34 & -0.01 & 0.22 & 0.97
& 0.10 & 0.11 & 0.24 & 0.13 & 0.22 & 0.98 \\
\addlinespace
$\sigma_g$ & 1.00 & 1.00 & 0.01 & 0.00 & 0.01 & 0.95
& 2.00 & 2.00 & 0.02 & 0.00 & 0.02 & 0.97
& 1.00 & 1.00 & 0.02 & 0.00 & 0.02 & 0.94 \\
$\omega_g$ & 1.50 & 1.51 & 0.08 & 0.01 & 0.07 & 0.97
& 1.00 & 1.03 & 0.05 & 0.03 & 0.06 & 0.90
& 2.00 & 2.03 & 0.10 & 0.02 & 0.11 & 0.93 \\
\addlinespace
$D_{g,11}$ & 1.00 & 1.02 & 0.07 & 0.02 & 0.08 & 0.93
& 1.00 & 0.99 & 0.12 & -0.01 & 0.12 & 0.93
& 1.00 & 0.99 & 0.08 & -0.01 & 0.08 & 0.97 \\
$D_{g,12}$ & 0.50 & 0.50 & 0.06 & -0.00 & 0.07 & 0.94
& 0.50 & 0.51 & 0.09 & 0.01 & 0.10 & 0.95
& 0.50 & 0.50 & 0.08 & -0.00 & 0.07 & 0.95 \\
$D_{g,22}$ & 1.00 & 0.98 & 0.10 & -0.02 & 0.09 & 0.96
& 1.00 & 1.01 & 0.13 & 0.01 & 0.14 & 0.95
& 1.00 & 1.01 & 0.12 & 0.01 & 0.12 & 0.94 \\
\bottomrule
\end{tabular}
\end{adjustbox}
\end{sidewaystable}

\begin{table}[ht]
\centering
\caption{Mean (in percentage) and standard deviation (in parentheses) of sensitivity, specificity, PPV, and NPV for the three classes in the simulation study with $G=3$.}
\label{sens_spec}
\begin{tabular}{c c c c c}
\toprule
Class & Sensitivity & Specificity & PPV & NPV \\
\midrule
\multirow{2}{*}{1} & 99.93\% & 99.97\% & 99.91\% & 99.97\% \\
& (0.04\%) & (0.02\%) & (0.05\%) & (0.02\%) \\
\addlinespace
\multirow{2}{*}{2} & 99.86\% & 99.87\% & 99.89\% & 99.85\% \\
& (0.05\%) & (0.05\%) & (0.05\%) & (0.05\%) \\
\addlinespace
\multirow{2}{*}{3} & 99.78\% & 99.94\% & 99.74\% & 99.95\% \\
& (0.11\%) & (0.03\%) & (0.11\%) & (0.03\%) \\
\bottomrule
\end{tabular}
\end{table}

\begin{table}[ht]
\centering
\caption{Confusion matrix of assigned versus true class for the simulation study with $G=3$. Percentages are shown with standard deviations in parentheses.}
\label{conf_matrix}
\begin{tabular}{lccc}
\toprule
\multirow{2}{*}{Assigned class} & \multicolumn{3}{c}{True class} \\
\cmidrule(lr){2-4}
& Class 1 & Class 2 & Class 3 \\
\midrule
Class 1 & 99.93\% (0.04\%) & 0.07\% (0.04\%) & $<0.01\%$ ($<0.01\%$) \\
Class 2 & 0.04\% (0.03\%) & 99.86\% (0.05\%) & 0.09\% (0.04\%) \\
Class 3 & $<0.01\%$ (0.01\%) & 0.22\% (0.11\%) & 99.78\% (0.11\%) \\
\bottomrule
\end{tabular}
\end{table}

\section{Application: Colorectal Cancer Dataset}
\subsection{Data description}

\begin{figure}[!htbp]
    \centering
\includegraphics[width=\textwidth]{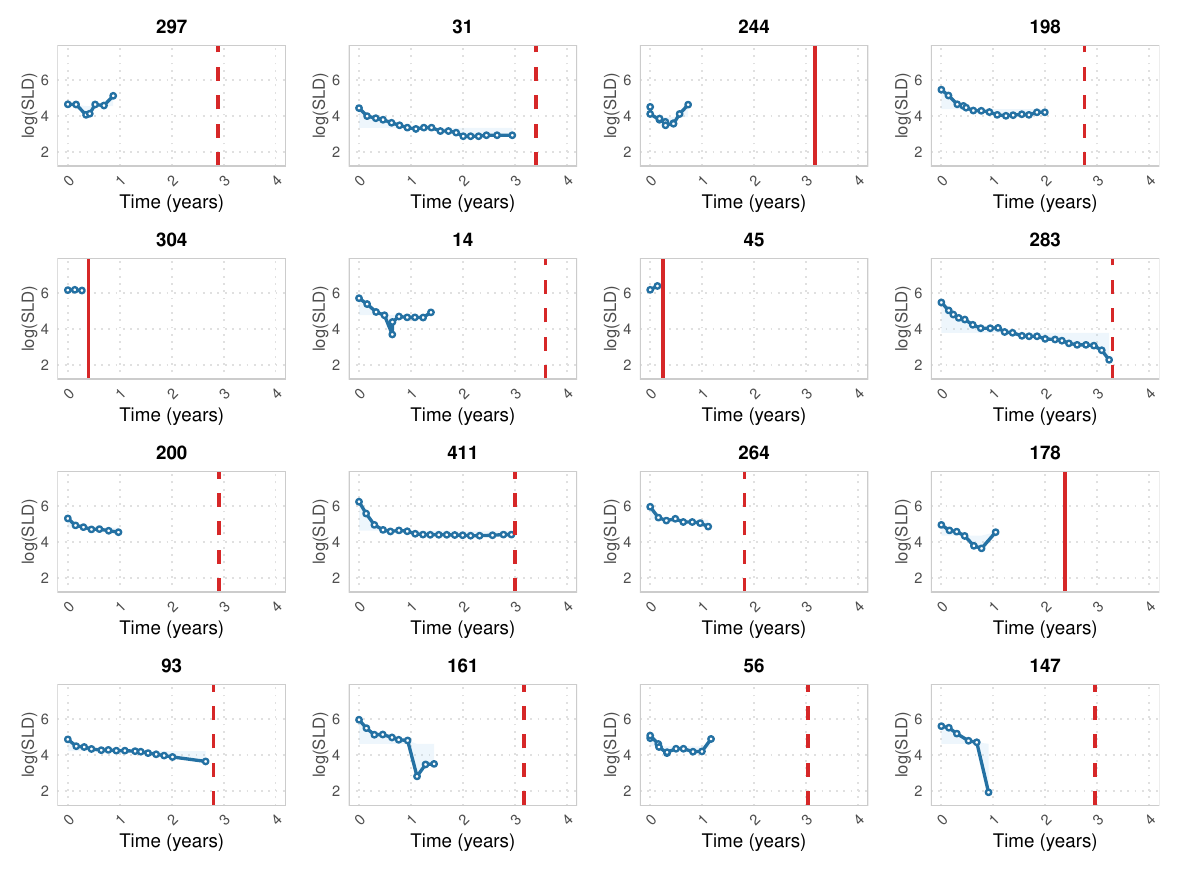}
    \caption{Log-transformed SLD trajectories for a randomly selected subset of patients from the colorectal cancer data, along with observed event time. Solid lines indicate an event, and dashed lines indicate censoring.}
    \label{panel}
\end{figure}

\begin{figure}[H]
    \centering
    \includegraphics[width=0.5\linewidth]{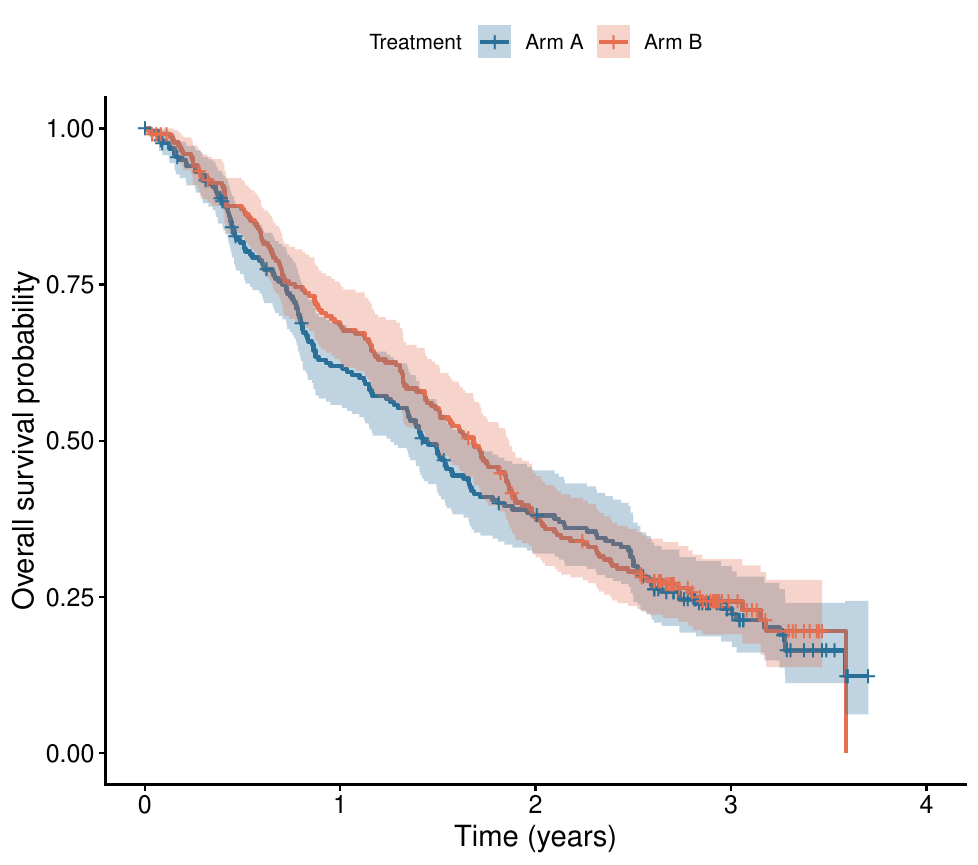}
 \caption{Kaplan-Meier survival curve for the colorectal cancer data stratified by treatment group. The blue line represent the patients in treatment arm A, and the orange line represents the patients in treatment arm B.}\label{km_real_data}
\end{figure}

Metastatic colorectal cancer refers to cancer that starts in the colon or rectum and spreads to other body parts. The disease causes growth of malignant tumours across the organism, and KRAS mutation is a driver in some cancers \cite{canellas2024mechanisms}. \\
PRIME is a multicentre international trial comparing the efficacy of treatment with FOLFOX4 alone and FOLFOX4 combined with panitumumab based on KRAS mutation in patients with metastatic colorectal cancer \cite{van2008open,normanno2009implications,bokemeyer2008kras}.. The trial consisted of $1183$ patients admitted in oncology clinics over the period August $2006$ to February $2008$. For our analysis, only $442$ patients were included in the study, with the rest omitted. We refer to this subset of the PRIME study as the colorectal cancer data. The longitudinal biomarker of interest is the sum of the longest diameter of target lesions (SLD). There are a total of $2266$ repeated measurements, with measurements of value $0$ excluded from the analysis. The mean (SD) number of measurements is $5.14$ ($3.45$), ranging between $1$ to $24$. The event of interest is time to death (in years), and patients who did not experience the event were censored. The mortality rate stands at 74\%, corresponding to 328 deaths.
\\
The patients were randomised into two treatment arms. Arm A represents the patients receiving treatment with FOLFOX4 alone and accounts for $50.5\%$ of the population, whereas arm B represents subjects receiving Panitumumab + FOLFOX4 and accounts for $49.5\%$ of the population.
Baseline covariates available in the data include sex and KRAS mutation status. $61.8\%$ of the patients were male, while $38.2\%$ were female. $58.8\%$ had Wild-type KRAS mutation, and $41.2\%$ were mutant.\\
Figure \ref{panel} displays individual logarithm of SLD longitudinal trajectories for $16$ randomly selected patients from the colorectal cancer data, with observed event times indicated by red coloured vertical lines. A solid line indicates an event, whereas a dashed line indicates censoring. As can be seen, some patients have steady trajectories, while others experience larger fluctuations in SLD values over time.
\\
Figure \ref{km_real_data} shows the Kaplan--Meier (KM) survival curve, illustrating the estimated survival probability over time stratified by treatment group for the colorectal cancer dataset. The plot suggests that there is no significant difference in survival probabilities between the two treatment groups.\\
The purpose of this section is to jointly analyse the longitudinal biomarker, namely the logarithm of SLD, and time to death in patients with metastatic colorectal cancer, by allowing class membership to evolve dynamically over time.

\subsection{Data analysis using the DJLCM}

In this section, we consider analysing the colorectal cancer dataset using the DJLCM with $G=2$ and $G=3$. We select the DJLCM with $G=2$ as the most suitable model. Results for the three-class DJLCM are reported in Appendix~\ref{djlcm_K3_analysis}. In addition, to investigate the contribution of the dynamic latent class structure and the shared random-effects association, we consider two simplified versions of the proposed DJLCM in which the latent class membership is static rather than time-varying. We refer to these models as the Bayesian standard JLCM (BJLCM), and Bayesian joint latent class model with shared random effects (BJLCM--SRE). The specifications of these comparative models, and full details about model selection are available in Appendix \ref{mod_selection}.

\paragraph{Class membership submodel}

Initially, each subject $i$ is classified into one of the $G$ latent classes,
$g \in \{1,\ldots,G\}$, with one of the classes taken as the reference class. The class membership probabilities are modelled using the
following multinomial logistic regression model:
\begin{equation*}\label{init_cls_app}
P\left(C_i(s_{i1})=g\right)
=
\frac{\exp\left(\delta_{g0}+\delta_{g1}\mathrm{SEX}_i\right)}
{\sum_{h=1}^{G-1}\exp\left(\delta_{h0}+\delta_{h1}\mathrm{SEX}_i\right)},
\qquad g=1,\ldots,G,
\end{equation*}
where $\mathrm{SEX}_i$ denotes the sex of subject $i$ (0 = female; 1 = male),
and $\delta_{r0}=\delta_{r1}=0$ for the reference class $r \in \{1, 2, \ldots, G\}$.
 For subsequent time points $s_{ij}$, $j = 2, \ldots, m_i$, the latent class process is assumed to follow a first-order discrete-time Markov model:
\begin{equation*}\label{cls_trans_app}
 P(C_i(s_{ij}) = l \mid C_i(s_{ij-1}) = g) 
= \frac{\exp(\eta_{gl0}+\eta_{gl1}\text{SEX}_i)}{\sum_{h=1}^{G} \exp(\eta_{hl0}+\eta_{hl1}\text{SEX}_i)}, 
\quad g,l = 1,\ldots,G,
\end{equation*}
where $\boldsymbol{\eta}_{gl}
=(\eta_{gl0},\eta_{gl1})^\top$ denotes the class-specific regression
coefficient vector governing the probability of transitioning from class
$g$ to class $l$, with $\boldsymbol{\eta}_{gr}=\mathbf{0}$ for all
$g=1,\ldots,G$, where $r \in \{1, 2, \ldots, G\}$ denotes the reference class.

\paragraph{Longitudinal submodel}
In this analysis,  the logarithm of the SLD trajectories is modelled via a class-specific linear mixed effects model, specifically 

\begin{equation*}
  \log(\text{SLD}_i(s_{ij})) \,\Big|\, _{C_i(s_{ij}) = g}
=\beta_{g0}+\beta_{g1}s_{ij}+\beta_{g2}\text{TRT}_i+\beta_{g3}\text{KRAS}_i+ b_{ig0}+b_{ig1}s_{ij}+\epsilon_{ijg},
 \end{equation*}
where $\mathrm{TRT}_i$ and $\mathrm{KRAS}_i$ denote the treatment group
(0 = FOLFOX4 alone; 1 = FOLFOX4 + panitumumab) and KRAS mutation status
(0 = wild-type; 1 = mutant) for individual $i$, respectively.
Here, $\bm{b}_{ig}=(b_{ig0}, b_{ig1})^\mathsf{T}\sim \mathcal{N}(0, \bm{D}_{g})$ are the subject-specific random intercept and slope and $\epsilon_{ijg}$ denotes the residual error, with
 $\epsilon_{ijg} \sim \mathcal{N}(0,\sigma_g^2).$

\paragraph{Survival submodel}



For the survival outcome, conditional on the latent class trajectory
$\bm{C}_i = \{C_i(s_{i1}),\ldots,$ $C_i(s_{im_i})\}$, which represents the
sequence of latent class memberships for subject $i$, and the
subject-specific random effects $\bm{b}_i$, where
$\bm{b}_i=(\bm{b}_{i1},\ldots,\bm{b}_{iG})^\top$ denotes the vector of
subject-specific random effects across the $G$ latent classes, the
hazard of the event for subject $i$ at time $t$ is modelled as
\begin{equation*}
\lambda_i(t \mid C_i(t)=g,\bm{b}_i)
=
\lambda_{0g}(t)
\exp\left\{
\gamma_{g0}
+\gamma_{g1}\mathrm{TRT}_i
+\gamma_{g2}\mathrm{KRAS}_i
+\alpha_{g0}b_{ig0}+\alpha_{g1}b_{ig1}
\right\},
\end{equation*}
where $\lambda_{0g}(t)$ denotes the class-specific Weibull baseline hazard,
specified as
$\lambda_{0g}(t)
=
\omega_{g}\,t^{\,\omega_{g}-1}$,
with $\omega_{g}$ denoting the class-specific Weibull parameter.
The coefficients $\gamma_{g0}$, $\gamma_{g1}$, and
$\gamma_{g2}$ are class-specific regression coefficients, while
 $\alpha_{g}=(\alpha_{g0}, \alpha_{g1})^\top$ is the vector of class-specific association parameters.\\
We use uninformative priors for Bayesian implementation and assume that
the model parameters are a priori independent. Specifically:

\begin{equation*}
\begin{alignedat}{2}
\delta_{g0},\ \delta_{g1} &\sim \mathcal{N}(0, 100), &\qquad & g \neq r,\\
\eta_{gl0},\ \eta_{gl1} &\sim \mathcal{N}(0, 100), && l \neq r,\\
\beta_{g0},\ \beta_{g1},\ \beta_{g3},\ \beta_{g4} &\sim \mathcal{N}(0, 100), &&\\
\gamma_{g0},\ \gamma_{g1},\ \gamma_{g2} &\sim \mathcal{N}(0, 100), &&\\
\alpha_{g0},\ \alpha_{g1} &\sim \mathcal{N}(0, 100), &&\\
\bm{L}_g &\sim \text{LKJ-Cholesky}(2), &&\\
\tau_{gj} &\sim \mathrm{half\text{-}Cauchy}(0, 5), && j = 1, 2,\ldots, G,\\
\sigma_g &\sim \mathrm{half\text{-}Cauchy}(0, 5), &&\\
\omega_g &\sim \mathrm{Gamma}(0.01, 0.01), &&
\end{alignedat}
\end{equation*}
where $g = 1, \dots, G$ unless stated otherwise. \\
The proposed DJLCM is implemented within a Bayesian framework using HMC  implemented in \texttt{Stan} via the \texttt{cmdstanr} package \cite{cmdstanr}. For implementation, we ran $80{,}000$ HMC iterations, including $40{,}000$ iterations as warm-up. The remaining $40{,}000$ iterations were used for posterior inference. \\

\subsubsection{Posterior parameter estimates} \label{djlcm_sel}

Table \ref{posterior_djlcm} reports the posterior summaries of the DJLCM with $G=2$. 
Observe that the association parameters $\alpha$ of both classes are significant, demonstrating a significant association between SLD levels and time to death.
For the longitudinal trajectories, SLD levels of class $2$ decrease over time ($\beta_{21}$), while the effect of time is insignificant for class $1$ ($\beta_{11}$).
The results also indicate no significant difference between treatment groups within either class ($\beta_{g2}$).
 Additionally, although positive, $\beta_{g3}$ indicates that KRAS mutation status does not significantly affect longitudinal trajectories within either class. Overall, this suggests there is not enough evidence to support the claim that SLD levels differ significantly according to KRAS status or treatment group.
 For the latent class process, we can see that sex is insignificant in both the initial class membership, and the transition model. 
\noindent
On the other hand, the results in the survival model report that KRAS mutation ($\gamma_{g2}$) does not have a significant effect on event risk in either class. Moreover, ($\gamma_{g1}$) indicates no significant difference in the effect of treatment arm on event risk. 
 \\ 
 Overall, the results suggest that the biggest driver for death in these cancer patients are the SLD levels. 

\noindent The results on covariate effects mainly agree with recent work on the colorectal cancer data. Indeed, Baghfalaki et al.~\cite{ganjali2026unified} analysed the data under a joint modelling framework and found that both treatment arm and KRAS mutation status were insignificant in the longitudinal and survival submodel. Rustand et al.~\cite{rustand2022two, rustand2023bayesian}, however, concluded that there was no significant difference in treatment effect in the survival submodel, while patients in treatment arm A were associated with a decrease in SLD levels over time. This reduction was worsened for mutant patients, or patients in treatment arm B. The earlier studies ~\cite{douillard2013panitumumab} suggest no significant difference in treatment effect for the risk of death in patients with mutant KRAS status, whereas treatment B was associated with a reduced risk in patients with Wild-Type KRAS mutation.  
\begin{table}[htbp]
\centering
\small
\setlength{\tabcolsep}{4pt}
\caption{Parameter estimates (Mean), standard deviation (SD), $2.5\%$ lower bound of $95\%$ credible interval, $97.5\%$ upper bound of $95\%$ credible interval and $\widehat{R}$: Gelman-Rubin diagnostic for analysing the Colorectal Cancer data using the DJLCM with $G=2$.}
\label{posterior_djlcm}
\resizebox{0.85\textwidth}{!}{%
\begin{tabular}{l@{\ }l rrrrr c rrrrr}
\toprule
 & & \multicolumn{5}{c}{Class 1 ($g=1$)} & & \multicolumn{5}{c}{Class 2 ($g=2$)} \\
\cmidrule(lr){3-7} \cmidrule(lr){9-13}
\multicolumn{2}{l}{Parameter} & Mean & SD & 2.5\% & 97.5\% & $\hat{R}$ & &
Mean & SD & 2.5\% & 97.5\% & $\hat{R}$ \\
\midrule
\multicolumn{13}{l}{\textit{Initial class membership submodel} } \\
Intercept     & ($\delta_{g0}$)  & -2.26  &  0.40 &  -3.15 &  -1.56 &  1.00 & & \multicolumn{5}{c}{Reference class} \\
\text{SEX}    & ($\delta_{g1}$)  & 0.51 &   0.51 &   -0.50 &   1.52 &  1.00 & & \\
\addlinespace
\multicolumn{13}{l}{\textit{Class transition submodel}} \\
Intercept     & ($\eta_{g10}$)   & 2.67 &   0.25 &   2.21 &   3.17 &  1.00 & & -1.41  &  0.16 &  -1.72 &  -1.11  & 1.00 \\
\text{SEX}    & ($\eta_{g11}$)   & 0.56 &   0.43 &   -0.26 &   1.44 &  1.00 & & -0.10 &   0.23 &   -0.55 &   0.34 &  1.00  \\
\addlinespace
\multicolumn{13}{l}{\textit{Longitudinal submodel}} \\
Intercept     & ($\beta_{g0}$)   & 4.75  &  0.119  &  4.52 &   4.98 &  1.00 & & 5.49 &    0.063 &   5.37  &   5.62 &  1.00 \\
Time          & ($\beta_{g1}$)   & -0.05 &  0.07 &  -0.19 &   0.09 & 1.00 & & -0.93 &  0.09 & -1.10 &  -0.76 &  1.00 \\
\text{TRT}    & ($\beta_{g2}$)   & -0.02 & 0.14 &   -0.30 &   0.26 &  1.00 & & -0.07 &  0.08 &  -0.22 &  0.09 & 1.00  \\
\text{KRAS}   & ($\beta_{g3}$)   &0.08 & 0.15 &   -0.20 &   0.37 &  1.00 & & $<$ 0.01 & 0.09  & -0.16 &   0.16 & 1.00 \\
$\sigma_{g}$  &                  & 0.10 &  0.01 & 0.09 &  0.11 &  1.00 & & 0.18 &  0.01 & 0.16 &   0.20 &  1.00 \\
$D_{g,11}$    &                  & 0.83 &  0.10 &   0.66 &   1.03 &  1.00 & & 0.52 &  0.04 &   0.44 &   0.61 &  1.00 \\
$D_{g,12}$    &                  & 0.05 &  0.05 &  -0.06 &  0.16 &  1.00 & & 0.03 &  0.05 &  -0.07 &  0.13 & 1.00 \\
$D_{g,22}$    &                  & 0.23 &   0.06 &   0.13 &  0.36 &  1.00 & & 0.69 &   0.13 &    0.45 &   0.97 &  1.00 \\
\addlinespace
\multicolumn{13}{l}{\textit{Survival submodel}} \\
$\omega_{g}$  &                  & 2.17 &   0.29 &    1.66 &   2.79  & 1.00 & & 1.46 &   0.10 &   1.27  &  1.67  & 1.00 \\
Intercept     & ($\gamma_{g0}$)  & -2.49  &  0.43 &   -3.39 &  -1.71 &  1.01 & & -1.12 &   0.19 &   -1.52 &  -0.77 &  1.00 \\
\text{TRT}    & ($\gamma_{g1}$)  & 0.06 &  0.26 &   -0.46 &   0.57 &  1.00 & & 0.23 &  0.19 &   -0.15 &   0.61 &  1.00 \\
\text{KRAS}   & ($\gamma_{g2}$)  & 0.45 &   0.26 &   -0.06 &  0.98 & 1.00 & & 0.26 &   0.20 &   -0.14 &   0.65 &  1.00 \\
$\alpha_{g0}$ &                  & 0.92 &   0.17 &    0.60 &   1.29 &   1.00 & & 0.61 &  0.14 &    0.34 &   0.90 &  1.00 \\
$\alpha_{g1}$ &                  & 0.84 &   0.43 &   -0.01 &  1.68  & 1.00 & & 1.04 &    0.20 &   0.66 &   1.45  & 1.00 \\
\bottomrule
\end{tabular}%
}
\end{table}
\FloatBarrier

\subsubsection{Estimated longitudinal trajectories} \label{djlcm_est_traj}
\begin{figure}[H]
    \centering
\includegraphics[width=1\linewidth]{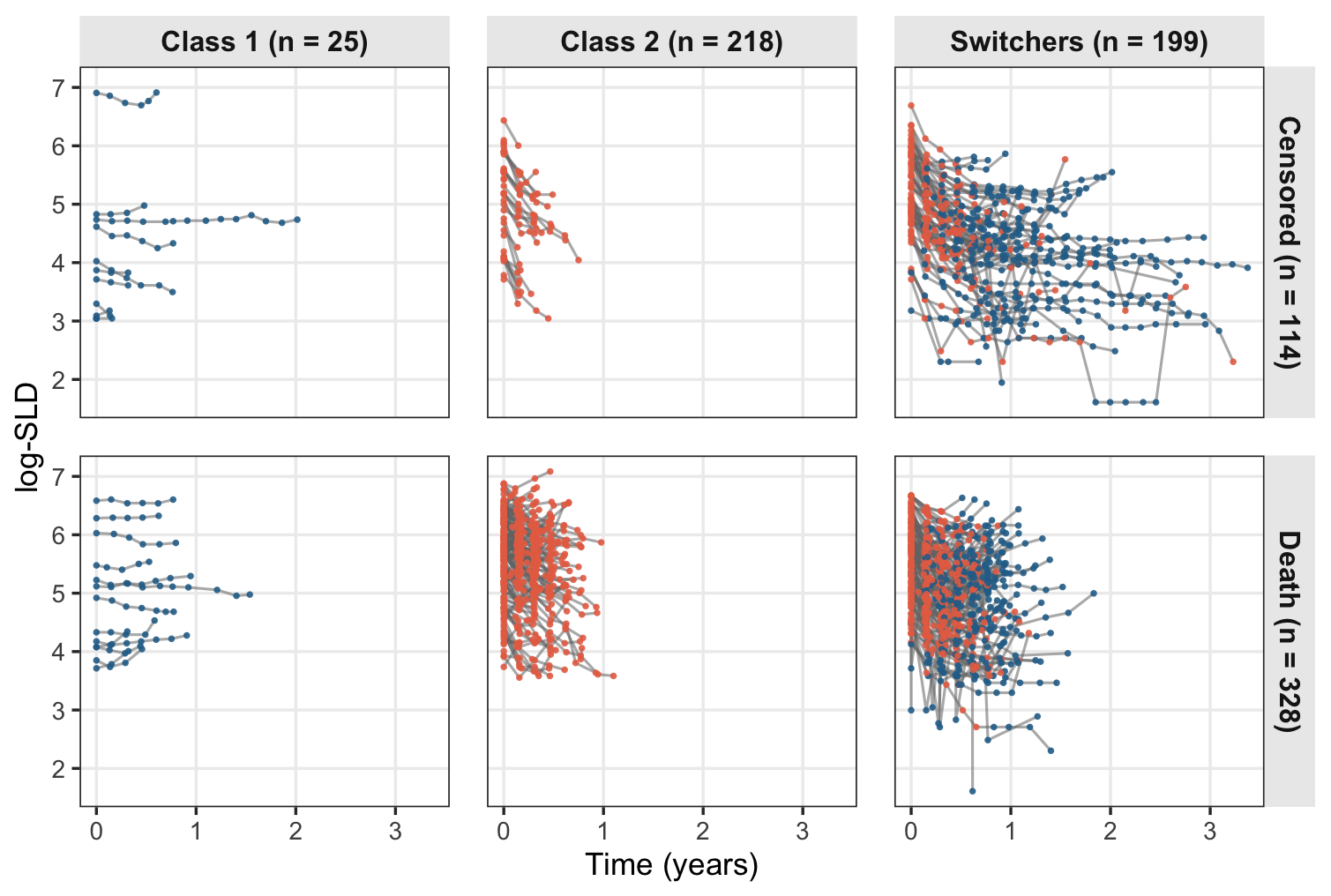}
    \caption{Longitudinal log-SLD trajectories with superimposed average trend lines, and grouped by class and event status for the DJLCM with $G=2$ for analysing the colorectal cancer data. Blue points indicate class $1$, and orange points class $2$. }
    \label{traj_class_djlcm}
\end{figure}

\noindent 
Figure \ref{traj_class_djlcm} reports the longitudinal SLD trajectories stratified by posterior class membership and event status. In our DJLCM, class $1$ accounts for $888$ ($39\%$) of repeated measurements, while class $2$ for $1378$ ($61\%$). $199$ ($45\%$) of individuals switch between classes during their follow-up, while $243$ ($55\%$) remain in one class. Among those remaining in one class, $25$ ($10\%$) are in class $1$, and $218$ ($90\%$) in class $2$. 
Out of patients in class $1$, $13$ ($52\%$) are on the treatment arm A, $15$ ($60\%$) are KRAS mutant, and event rate is $52\%$. In class $2$, $108$ ($50\%$) are on treatment arm A, $90$ ($41\%$) are mutant, and event rate is $83\%$.
Among the patients who switch between latent classes, $103$ ($52\%)$ are on treatment group A, $77$ ($39\%)$ are mutant, whereas event rate is $67\%$. We refer to the individuals who switch between classes as switchers.

\noindent Observe from the middle plot that patients classified in Class $2$ are characterised by a steeply decreasing trajectory, indicating shrinking tumour size.
From the left graph, 
patients who remain in class $1$ have a flat, stable trajectory, indicating steady tumour levels. These class trajectories are in line with the results from Table \ref{posterior_djlcm} regarding the effect of time on SLD levels. The third plot indicates that among patients who switch between classes, most of them jump from class $2$ to class $1$.
Moreover, for switchers, the patients who experience the event have shorter trajectories compared to those event-free, but this is not necessarily the case in class $1$ or $2$.

\begin{figure}[H]
\centering
\includegraphics[width=0.7\linewidth]{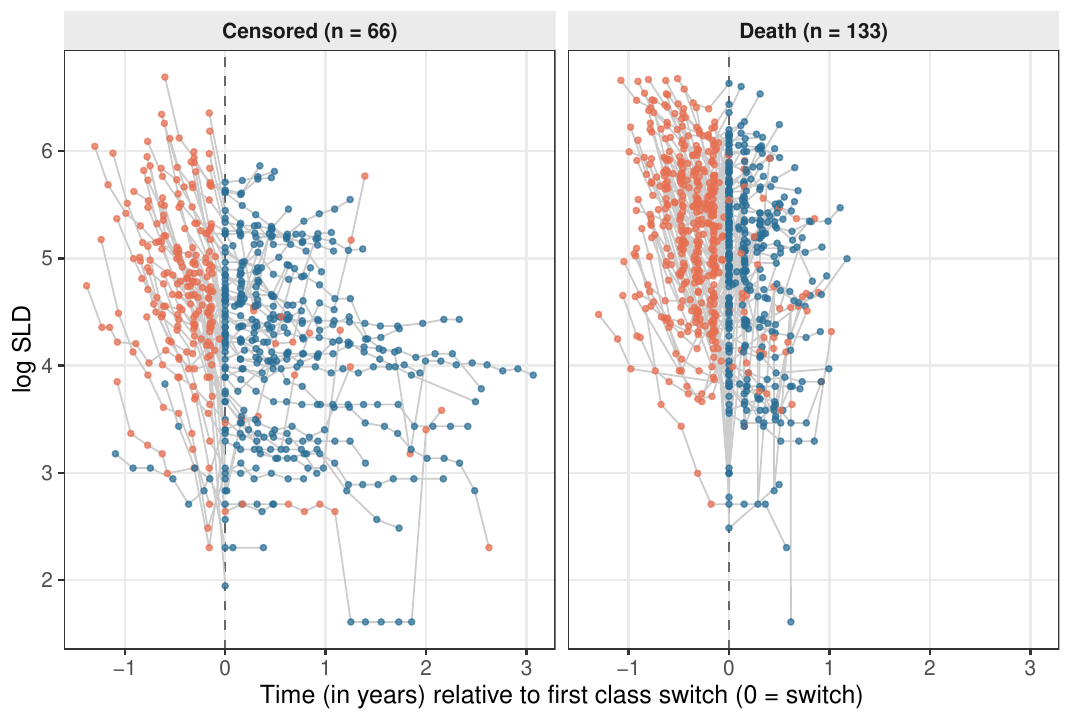}
    \caption{Longitudinal trajectory relative to first switch time for the DJLCM with $G=2$ for analysing the colorectal cancer data. Blue points indicate class $1$, and orange points class $2$.}
    \label{switch_rs}
\end{figure}

\noindent To better understand what the switch represents, Figure \ref{switch_rs} presents the SLD levels of the $199$ subjects who switch between classes, with the time of the first switch as the origin. Among those patients $171$ switch from class $2$ to class $1$, only $1$ patient moves from class $1$ to $2$, and $27$ patients have multiple switches. The mean time at first switch is $0.503$. The figure illustrates how tumour size behaves before and after a switch. A switch is characterised by a change in the trajectory pattern: from steeply decreasing to flat for patients who do not experience the event (left), and from sharply decreasing to increasing for most patients who experience the event (right). The right plot indicates that an abrupt change is a more significant signal, as these patients experienced the event. The first switch, can therefore be seen as an important turning point in a patient's tumour size. In retrospect, such a switch could necessitate a change in therapy, or more intensive monitoring of the patient's health condition.
\\

\noindent In summary, although the individual treatment effects cannot be captured from the available information, the decreasing trend of the longitudinal trajectories suggests that treatment is having a positive effect in reducing tumour size for these patients. However, the results also indicate that there is no significant difference between treatment arm in either class. The analysis further suggests that the subjects classified consistently in class $1$ could have a steady, controlled tumour burden, while those in class $2$ may be experiencing a more aggressive cancer with fast-dividing cells. Although treatment reduces the tumour size in these patients, due to their significantly higher tumour size, it doesn't seem to be a significant factor in the survival rate, with $83\%$ of them experiencing death.
Moreover, due to the dynamic class membership, our model identified a group of patients (switchers) for whom tumour size initially reduces significantly, but as time progresses, this trend fades and tumour size levels off, as indicated in the rightmost panel of Figure~\ref{traj_class_djlcm}. \\

\noindent Regarding the DJLCM with $G=3$, the estimated longitudinal trajectories and trajectories relative to first switch time for are presented in Figure \ref{traj_class_djlcm_k3}, and Figure \ref{switch}, respectively. For comparison between the two-class DJLCM and three-class DJLCM details are provided in Table \ref{model-comparison} and Table \ref{posterior_djlcm_g3}.

\noindent Comparing Figure \ref{traj_class_djlcm} and Figure \ref{traj_class_djlcm_k3}, observe that Class $1$ and $2$ from the three-class DJLCM mainly correspond to the classes in the DJLCM with $G=2$. We computed classification agreement between the two models using
the Adjusted Rand Index (ARI)~\cite{rand1971objective}, via the R package \texttt{aricode}~\cite{aricode}. The ARI measures the agreement between two clustering methods, and is reported to be $0.66$, which demonstrates good classification agreement between the DJLCM with $G=2$ and $G=3$.
\noindent Class $3$ in the case $G=3$ captures a group of patients with an increasing trajectory trend, although the number of such patients is only $7$. Looking at the plot for the switchers, the additional class is mainly a substructure of the switchers from the two-class DJLCM.
The three-class DJLCM correctly captures a tendency of the tumour burden to increase, however Figure \ref{switch} demonstrates that after that slight increase, tumour levels quickly level off for the majority of the patients. Hence, although correctly capturing a different trajectory pattern, the three-class DJLCM does not provide a clinically meaningful advantage over the two-class DJLCM for this particular dataset. 
Importantly, although the two-class DJLCM contains only two latent classes, it can capture more than two distinct patterns of class membership over time because subjects may remain in one class or transition between the two classes. 

\noindent Overall, due to the high classification agreement between the two-class and the three-class DJLCM, and considering the much higher number of parameters for the three-class DJLCM, we decided to retain DJLCM with $G=2$ for computational efficiency and parsimony.


\subsection{Comparison with BJLCM and BJLCM--SRE} \label{comp}

Table~\ref{model-comparison} reports the comparison of BJLCM, BJLCM--SRE, and DJLCM in terms of LPML, WAIC and BIC. 


\noindent The BJLCM was fitted for $G=1,2,3,4$ classes. In the case $G=1$, the BJLCM turns into a model with independent longitudinal and survival components. 
Although the criteria favour BJLCM with $G=4$, our analysis shows that class $2$ only captures $25$ ($5\%$) of the population and thus may not represent a clinically meaningful subgroup. Moreover, it would not be feasible to fit a BJLCM with $G=5$ classes to the colorectal cancer data due to the high missingness rate; the number of patients with less than $5$ repeated measurements is $224$ ($51\%)$. Therefore, in line with Weller et al.~\cite{weller2020latent}, and for computational feasibility, we decided to focus on the BJLCM with $G=3$. The results for the BJLCM with $G=3$ can be found in Table \ref{posterior_model3}.

\noindent The BJLCM--SRE was fitted for $G=1,2$ and $3$ classes. For $G=1$, the BJLCM--SRE becomes the conventional joint model with shared random effects~\cite{tsiatis1995modeling}. Again, the criteria favour the model with $G=3$. The results for the BJLCM--SRE with $G=3$ are presented in Table \ref{posterior_model3_alpha}.

\begin{figure}[htbp]
\centering
\begin{subfigure}[t]{0.48\textwidth}
  \centering
  \includegraphics[width=0.83\linewidth]{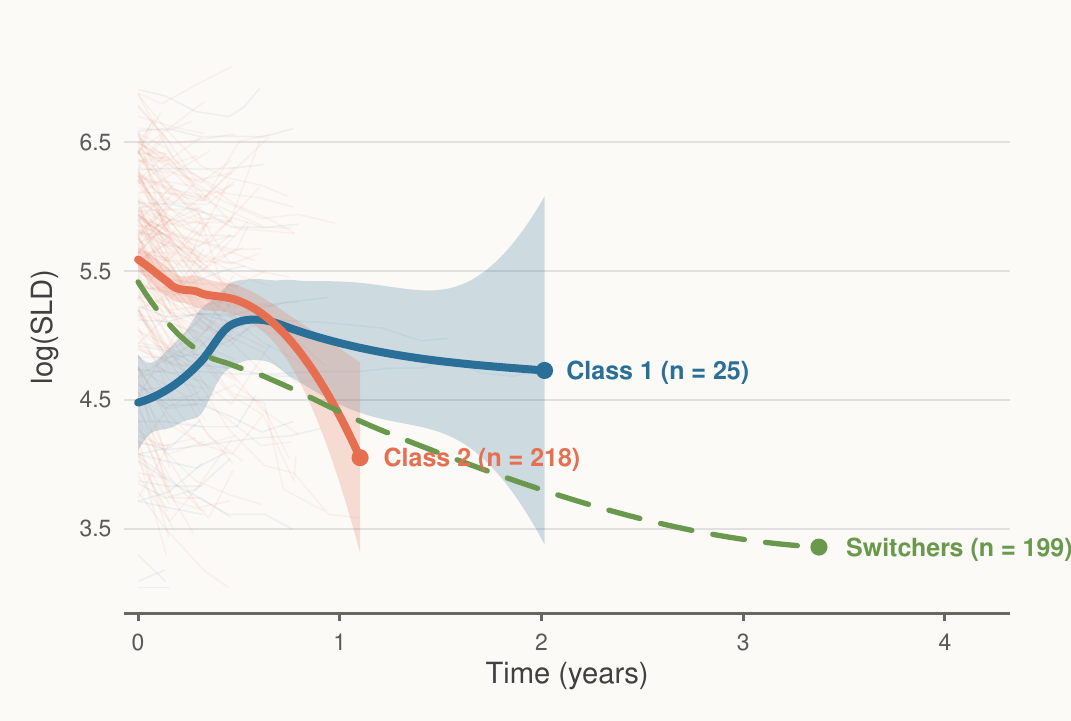}
  \caption{DJLCM ($G=2$)}
  \label{fig:traj-djlcm}
\end{subfigure}%
\hfill
\begin{subfigure}[t]{0.40\textwidth}
  \centering
  \includegraphics[width=\linewidth]{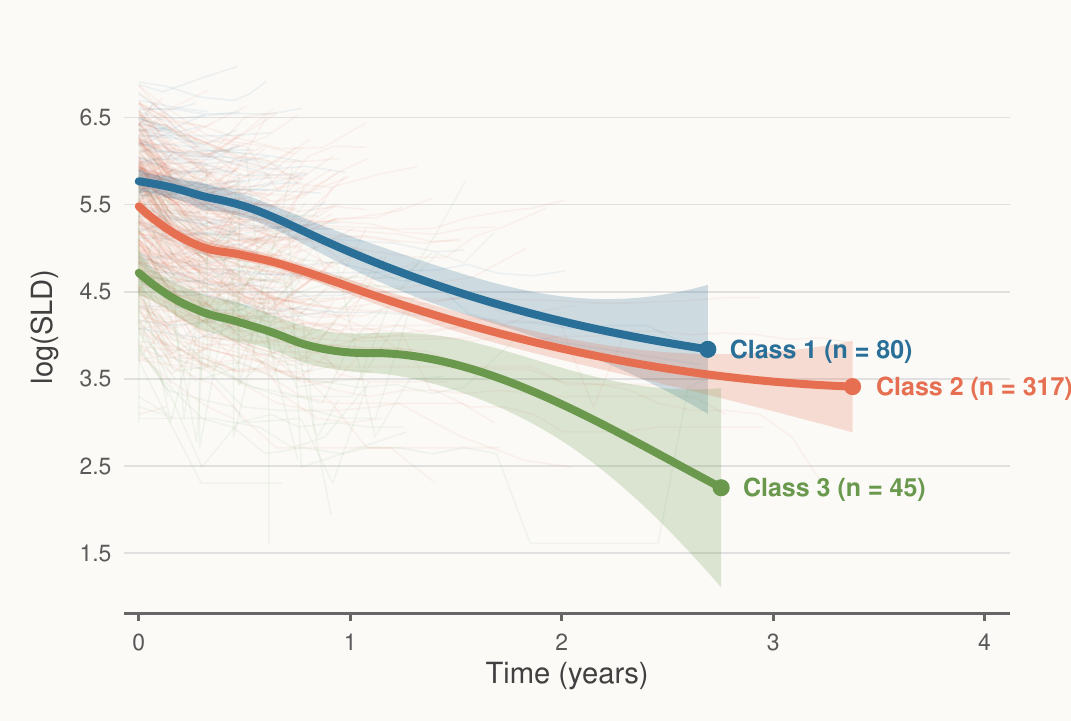}
  \caption{BJLCM ($G=3$)}
  \label{fig:traj-std-jlcm}
\end{subfigure}
\par\vspace{1.5em}
\begin{subfigure}[t]{0.40\textwidth}
  \centering
  \includegraphics[width=\linewidth]{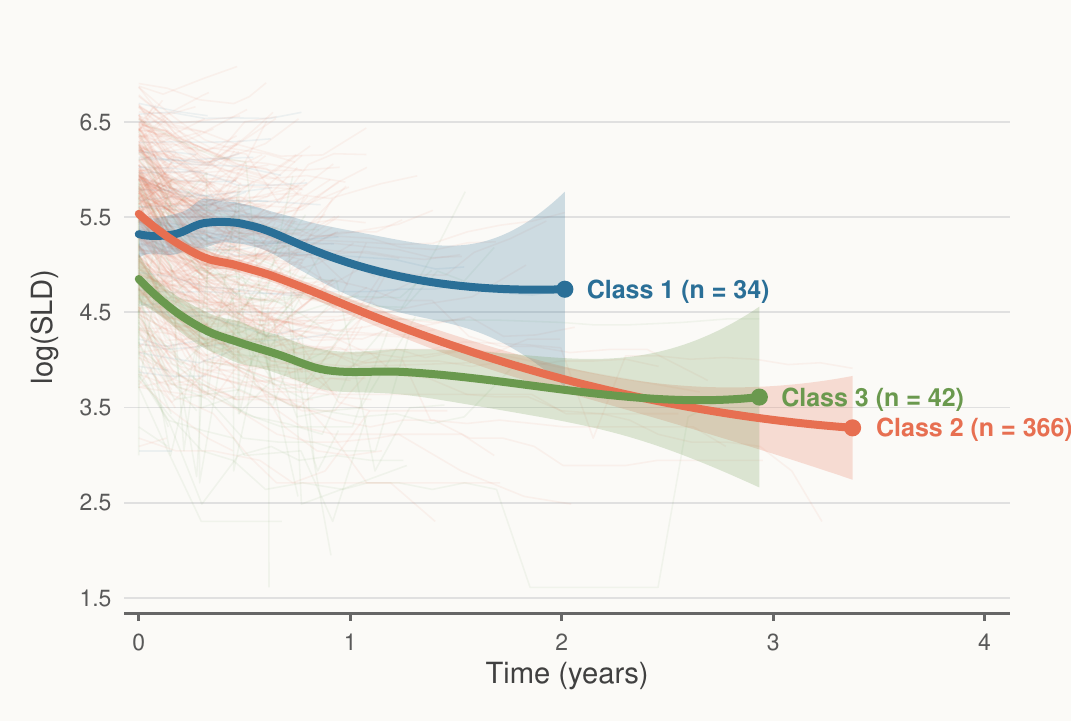}
  \caption{BJLCM--SRE ($G=3$)}
  \label{fig:traj-jlcm-shared}
\end{subfigure}
\caption{Class-specific mean longitudinal trajectories for the DJCLM with $G=2$ (panel a), BJLCM--SRE with $G=3$ (panel b), and the BJLCM with $G=3$ (panel c) for analysing the colorectal cancer dataset.}
\label{fig:4_models}
\end{figure}

\noindent Figure \ref{fig:4_models} illustrates the class-specific non-parametric mean longitudinal trajectories for the BJLCM ($G=3)$, BJLCM--SRE ($G=3)$, and DJLCM ($G=2)$. Figure~\ref{fig:traj-djlcm} reports the mean longitudinal trajectories for each class (based on the individuals remaining in that class throughout the whole study) in the DJLCM, along with the mean for the switchers.
The baseline log(SLD) levels for  subjects who start in class $1$ are $4.36$, while for class $2$ they are significantly higher at $5.53$, with the Wilcoxon rank-sum test \cite{wilcoxon1945individual} for testing the significance of the difference in means of baseline SLD levels between the two classes reporting a statistic $W = 2262$ and a $p$-value $p < 0.001$. Patients who are consistently in class $2$ have sharply decreasing and shorter trajectories compared to those in class $1$, who have stable trajectories. Meanwhile, patients who switch have the longest and a steadily decreasing trajectory. The mean number of repeated measurements for class $1$ is $4.88$, for class $2$ is $3.14$, and $7.34$ for the switchers. Observe from Figure \ref{fig:traj-std-jlcm} that the BJLCM ($G=3$), mainly classifies the subjects according to the magnitude of their SLD levels, with all classes having similar slopes. In contrast, our DJLCM classifies mainly based on trajectory shape, allowing for deeper insights into changes of tumour size course. On the other hand, comparing the BJLCM--SRE (Figure~\ref{fig:traj-jlcm-shared}) with $G=3$ to our DJLCM, we can see that class $2$ of the BJLCM--SRE mostly aligns with the switchers in the DJLCM, with $156$ subjects in common. Class $1$ of the DJLCM captures some subjects from Class $1$ and Class $3$ in Figure \ref{fig:traj-jlcm-shared}. This reinforces our model's ability to capture more than $2$ classes due to the dynamic class membership.

\paragraph{Comparison of survival submodels}

On the other hand, Figure~\ref{fig:4_models_surv} presents the survival curves for the BJLCM ($G=3$), and the BJLCM--SRE $(G=3)$ from the analysis of the colorectal cancer data. Figure~\ref{fig:a_surv} suggests that for the BJLCM with $G=3$ classes, patients in Class $1$ have consistently lower survival probabilities compared to those in other classes. For the BJLCM--SRE with $G=3$ classes, Figure~\ref{fig:b_surv} indicates that class $3$ have consistently better overall survival probabilities compared to the other two classes, while $1$ and $2$ have similar survival curves. 

\begin{figure}
    \centering    \includegraphics[width=0.8\linewidth]{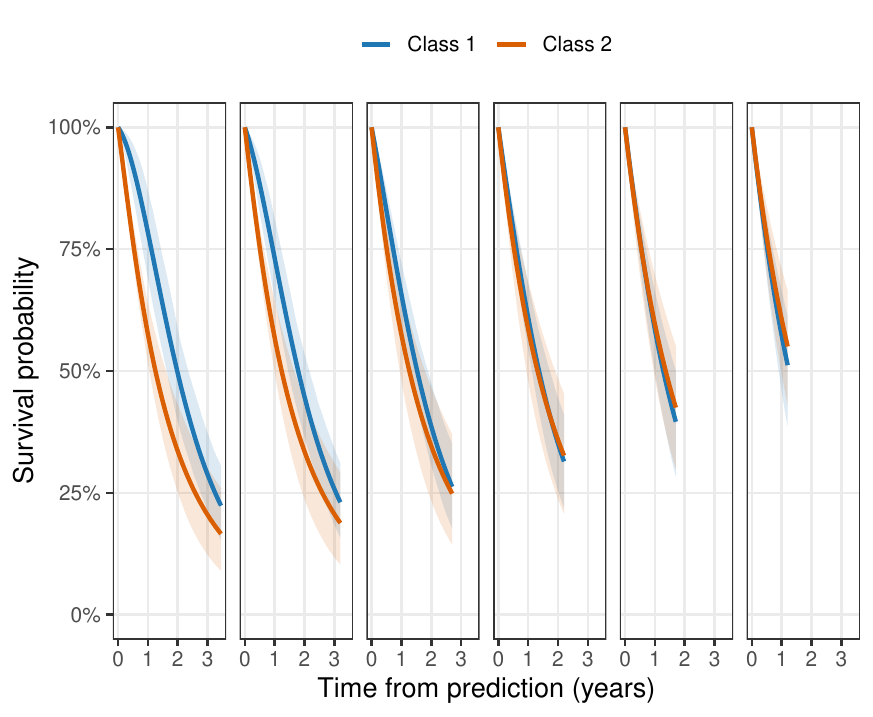}
    \caption{Class-specific survival curves for the DJLCM ($G=2)$ with reference times 0.25, 0.5, 1, 1.5, 2, and 2.5, left to right, for analysing the colorectal cancer dataset.}
    \label{surv_curves_analysis}
\end{figure}

\noindent Due to the dynamic class membership of the DJLCM, traditional KM curves are not appropriate to use because the risk set for each class is determined at the start of the follow-up and assumed constant. 
Therefore,
to evaluate the effect of class membership on the survival prognostics, we estimated class-specific survival curves for the DJLCM with $G=2$ by fitting model-based survival curves at different times during the follow-up period. At each reference time $s$, patients still at risk were classified according to their most recent posterior class assignment at or before $s$. Class was kept fixed from that time on, and separate Cox models~\cite{van2007dynamic} adjusted for the effects of treatment arm and KRAS mutation status were fitted to the subjects in each class. The patients were followed for four years from the reference time. This allowed us to better assess how the risk differs between the two classes in the DJLCM. 
Figure~\ref{surv_curves_analysis} reports the estimated class-specific survival curves at times $0.25, 0.5, 1$, $1.5$, $2$, and $2.5$ years. 
In all but the last two cases, patients in class $1$ seem to have better survival probability compared to those in class $2$. At prediction times $0.25$ and $0.5$, class $2$ has significantly lower survival probability than class $1$, as indicated by the non-overlapping confidence bands of the two curves.
However, the difference in survival probabilities between the two classes becomes insignificant after time $1$, indicating that the $1$ year window is important because it represents a change in the prognostic effect of class membership. As mentioned before, the mean time at first switch is $0.503$. On the other hand, the mean event time for patients remaining in class $1$ ($2$) is $1.59$ ($0.98)$, respectively. This suggests that by time $1$ or $1.5$, most of the patients who remain in class $2$ would have experienced the event, and the majority of others would have already moved to class $1$. So the insignificant difference in survival probability could be due to the fact that those surviving the 1 year mark, and staying in class $2$ later could genuinely have better survival chances after the initial reactive period, or it could be that the survival probability for class $1$ is being affected by patients who transited from class $2$ to class $1$. 
Either way, the analysis suggests that class membership is an important prognostic factor in the earlier stages of follow-up , within $1$ year, and the effect fades for patients surviving beyond that point.
\\
In summary, the BJLCM $(G=3)$ and BJLCM--SRE ($G=3$) predict class $1$ to have constantly higher risk of event, followed by Class $2$, and then Class $3$. In contrary, our DJLCM predicts the risk of event between the classes to be significantly different only up to $1$ year follow-up time. Hence, the DJLCM is able to provide more meaningful clinical insights into the prognostic power of class membership.

\section{Discussion}

Joint models are widely used to analyse longitudinal and survival data by considering the association between them.
The standard JLCM~\cite{lin2002latent, proust2014joint} was introduced as an alternative to the shared random effects model to allow for heterogeneous populations through the latent classes. In this paper, we proposed a new, flexible dynamic joint latent class model with time-varying class membership. The standard JLCM, along with its extensions proposed in the literature~\cite{andrinopoulou2020integrating, liu2015joint, proust2014joint}, assume that the class membership is fixed, hence subjects remain in one class during their entire follow-up. We relax this assumption by allowing the subjects to switch classes at discrete measurement timepoints. The dynamic class membership process is governed by a first-order discrete Hidden Markov Model. A Bayesian approach using Hamiltonian Monte Carlo (HMC) was adopted for inference. The implementation was carried out in the \texttt{cmdstanr}~\cite{cmdstanr} package in R. Since Stan cannot sample discrete variables~\cite{stan_users_guide}, such as latent class membership, we derived the likelihood by summing over all possible class paths for each subject. To reduce computational burden, we used the forward algorithm~\cite{baum1970maximization, rabiner1989tutorial}, which computes the class path at each timepoint recursively.
\\
We applied our model to the colorectal cancer data, which studies the effect of treatments and KRAS mutation status on patients with metastatic colorectal cancer. We jointly analysed the longitudinal SLD levels with time to death, while allowing for dynamic latent classes in the population. The findings suggest that the population consists of two classes, of which class $2$ is characterised by higher tumour sizes with a sharply decreasing longitudinal trajectory, and class $1$ with steady tumour levels over time. We also assessed the probability of survival within each class.
The findings report that the probability of experiencing the event between the two classes differed only in the earlier phases of follow-up, with no significant difference in survival probability after $1$ year. Moreover,  
the results indicate no significant effects of KRAS mutation status or treatment difference in either class, both on the longitudinal SLD levels and event risk.
\\
The DJLCM was compared to two JLCMs in the literature~\cite{proust2014joint, liu2015joint}, and demonstrated substantially better performance based on WAIC~\cite{watanabe2010waic}, LPML~\cite{gelfand1994cpo}, and BIC~\cite{schwarz1978estimating}.
The results indicate that the DJLCM provided a different classification perspective on the data compared to the standard JLCMs. Indeed, our model was able to capture fluctuations in trajectory patterns and detect changes in disease course through the jumping behaviour. Moreover, through the dynamic class process,  the DJLCM could potentially capture more than two classes. Overall, our model could identify a group of patients who start with high SLD levels decreasing sharply, and after some time, their SLD levels level off. This change is reflected in the switch between classes, providing us with a deeper understanding of trajectory patterns than the BJLCM, where patients remain in one class through the entire follow-up period.

\noindent Additionally, we conducted extensive simulation studies to validate our methodology. The model produced consistently low RMSE and bias, indicating precise estimation and good convergence. Furthermore, the misclassification analyses demonstrated excellent discriminative power of the proposed model. 
\\
However, there are limitations to the model. In our experience, in datasets where the event rate is very low, the model may struggle to estimate the parameters of a class with a low number of events, resulting in high uncertainty around them. In these cases, the model is sensitive to initial values, and different initial values may lead to different estimates for the particular class. Hence, the DJLCM can be best applied to datasets with higher event rates to ensure convergence. 

 \noindent Also, this work can be extended in several directions.
The DJLCM features the dynamic class membership which could be very attractive in datasets with longer follow-up and larger number of repeated measurements. However, as mentioned above, the current estimation approach involves summing over all possible class sequences at each timepoint for all subjects. This could prove computationally expensive for longitudinal data involving multiple biomarkers. Hence, another estimation approach, which allows for direct sampling of class membership, could be explored for more efficient inference.
Moreover, in the survival analysis conducted in Section \ref{comp} the survival curves are computed assuming fixed class membership from the reference point onwards, which ignores the dynamic nature of the latent class membership.  Also, the analysis does not utilise repeated SLD measurements, but rather relies only on the survival covariates for estimating the class-specific survival curves. Therefore, the approach presented here can be extended to dynamic risk prediction of survival probability by incorporating all available information from both the longitudinal and survival process up to landmark time~\cite{rizopoulos2011dynamic, proust2009development}. This would enable us to better assess the effect of the dynamic latent class process on the survival outcome, as well as obtain dynamic predictions of longitudinal trajectories. Furthermore, for the colorectal cancer data, the model selection criteria appear to favour models with larger number of classes. As a future work, developing more robust criteria for determining the optimal number of classes will be useful.


\subsection*{Acknowledgements}
The authors would like to acknowledge the assistance provided by Research IT and the use of the Computational Shared Facility at The University of Manchester.

 \bibliography{sample}

\newpage

\appendix
\section{Longitudinal trajectories for the simulation studies}
\setcounter{table}{0}
\renewcommand{\thetable}{A.\arabic{table}}
\setcounter{figure}{0}
\renewcommand{\thefigure}{A.\arabic{figure}}

\begin{figure}[H]
    \centering
    \begin{subfigure}{0.35\textwidth}
        \centering
        \includegraphics[width=\linewidth]{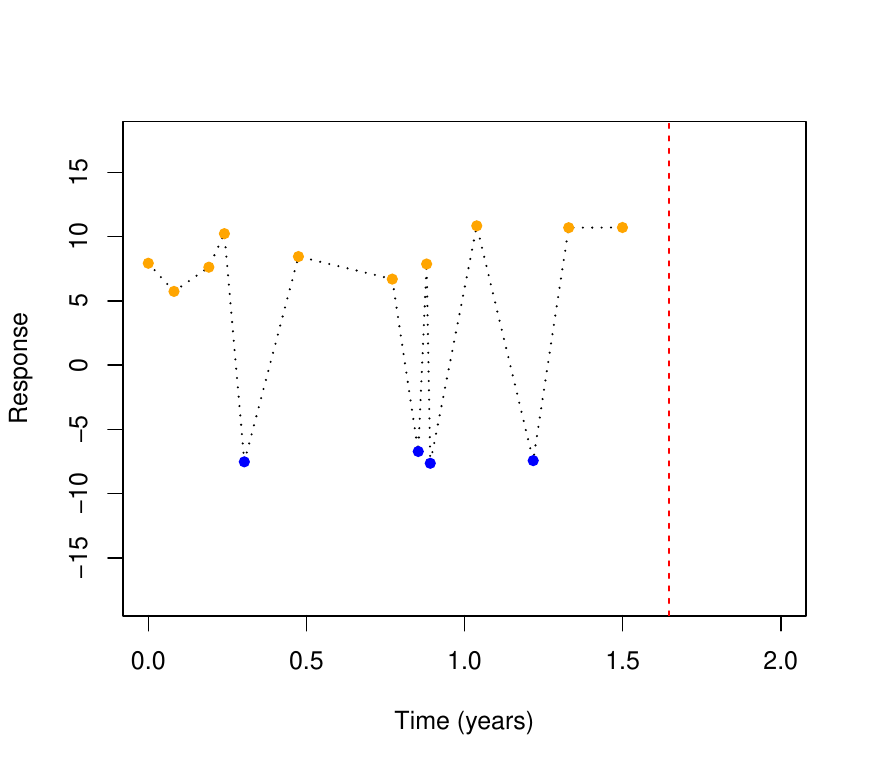}
        \caption{Subject $974$}
    \end{subfigure}
    \hfill
    \begin{subfigure}{0.35\textwidth}
        \centering
        \includegraphics[width=\linewidth]{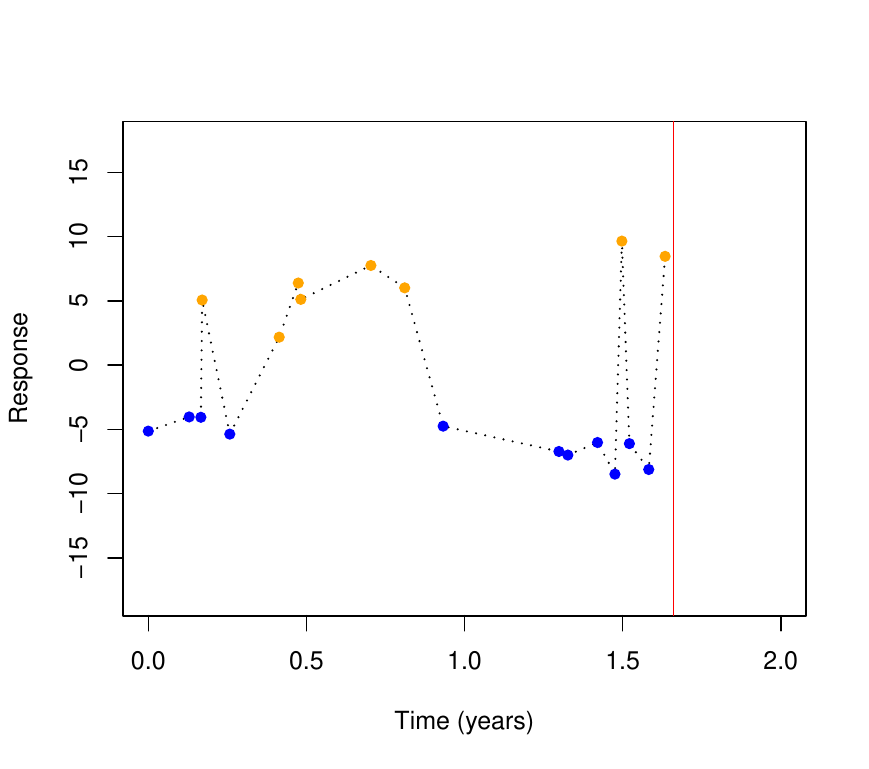}
        \caption{Subject $398$}
    \end{subfigure}
    \vspace{0.5cm}
    \begin{subfigure}{0.35\textwidth}
        \centering
        \includegraphics[width=\linewidth]{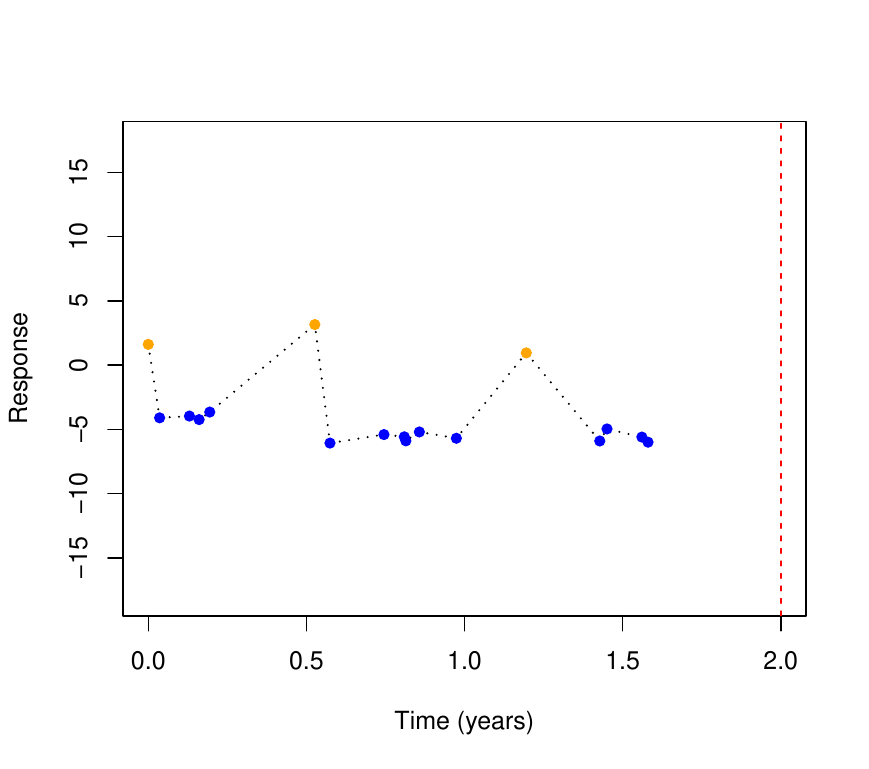}
        \caption{Subject $719$}
    \end{subfigure}
    \hfill
    \begin{subfigure}{0.35\textwidth}
        \centering
        \includegraphics[width=\linewidth]{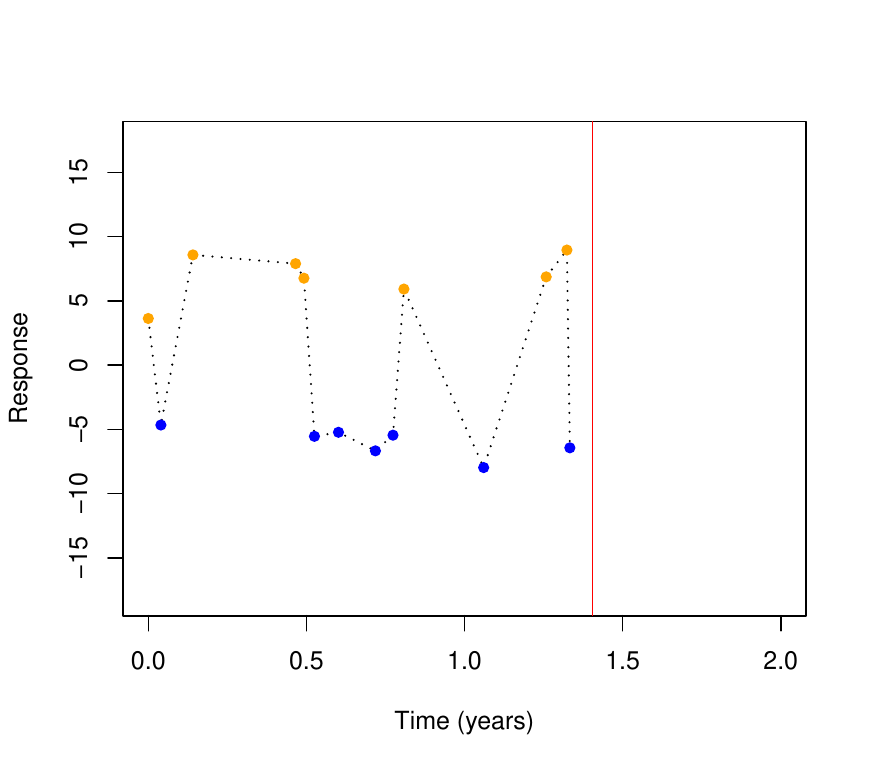}
        \caption{Subject $15$}
    \end{subfigure}
    \vspace{0.5cm}
    \begin{subfigure}{0.35\textwidth}
        \centering
        \includegraphics[width=\linewidth]{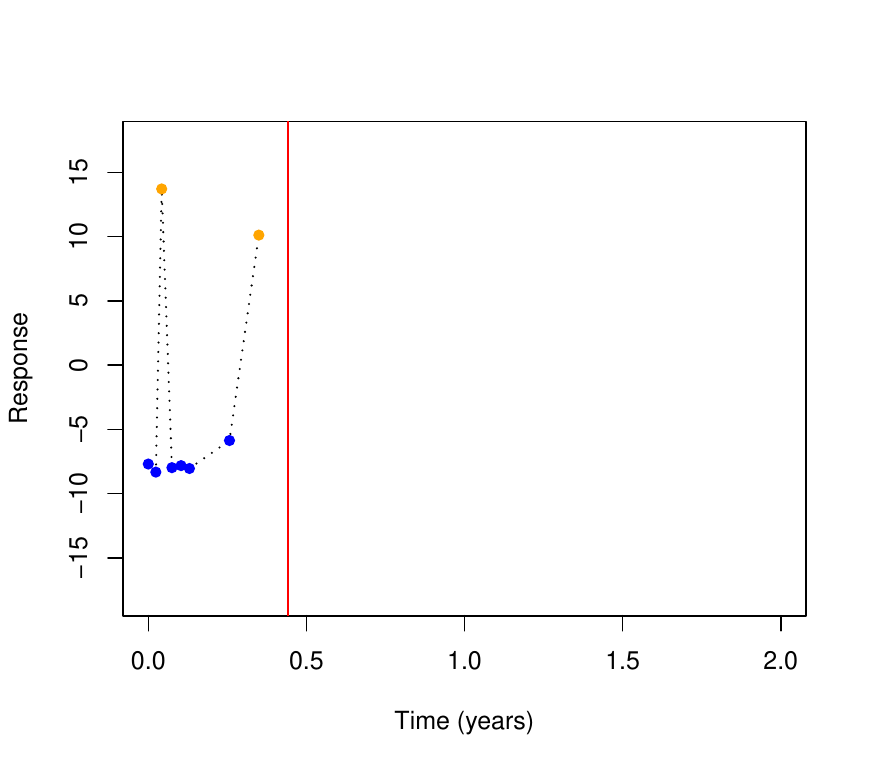}
        \caption{Subject $521$}
    \end{subfigure}
\caption{Individual longitudinal trajectories for five  randomly selected subjects from for simulation study with $G=2$, with observed event times indicated by a red line. Blue points represent class $1$, orange points class $2$, respectively. A solid line indicates an event, whereas a dashed line indicates censoring.}
\label{individual_trajectories_K2}  
\end{figure}

\begin{figure}[H]
    \centering
    \begin{subfigure}{0.40\textwidth}
        \centering
        \includegraphics[width=\linewidth]{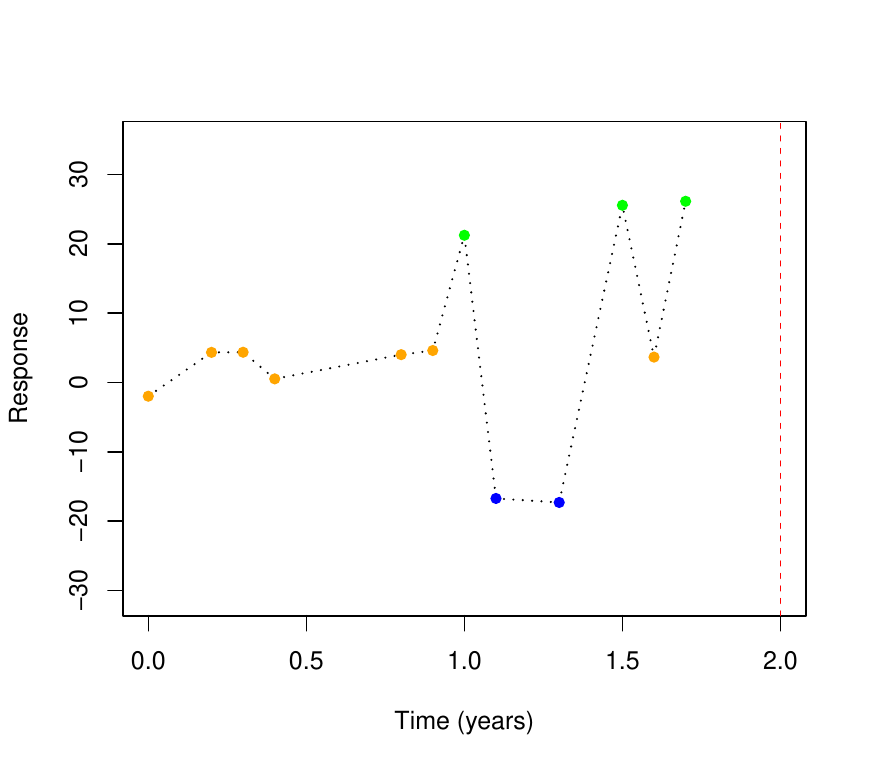}
        \caption{Subject 276}
    \end{subfigure}
    \hfill
    \begin{subfigure}{0.40\textwidth}
        \centering
        \includegraphics[width=\linewidth]{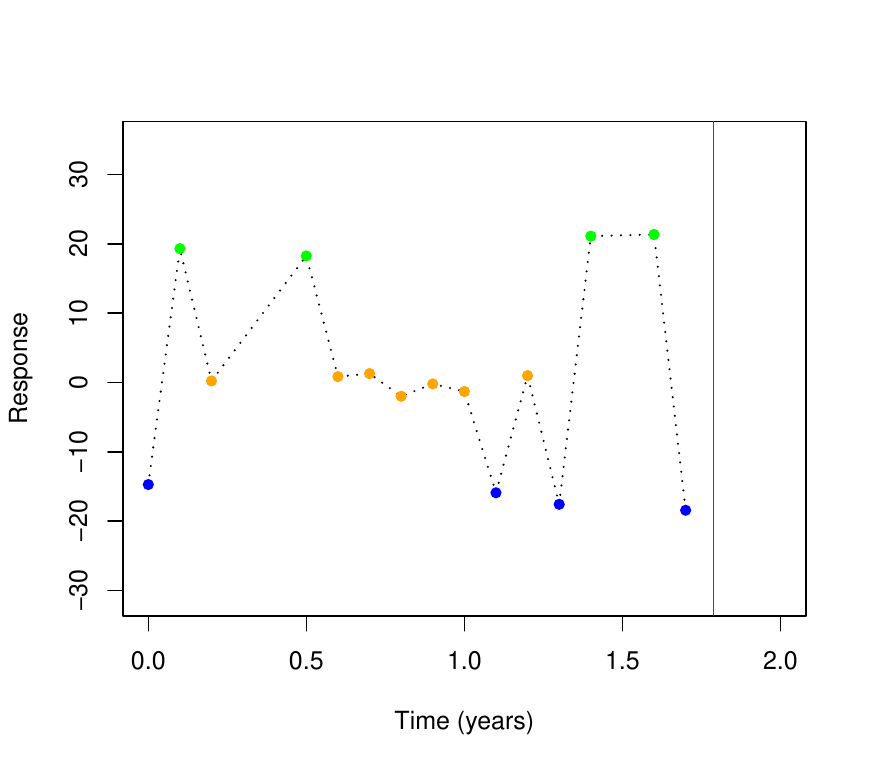}
        \caption{Subject 303}
    \end{subfigure}
    \vspace{0.5cm}
    \begin{subfigure}{0.40\textwidth}
        \centering
        \includegraphics[width=\linewidth]{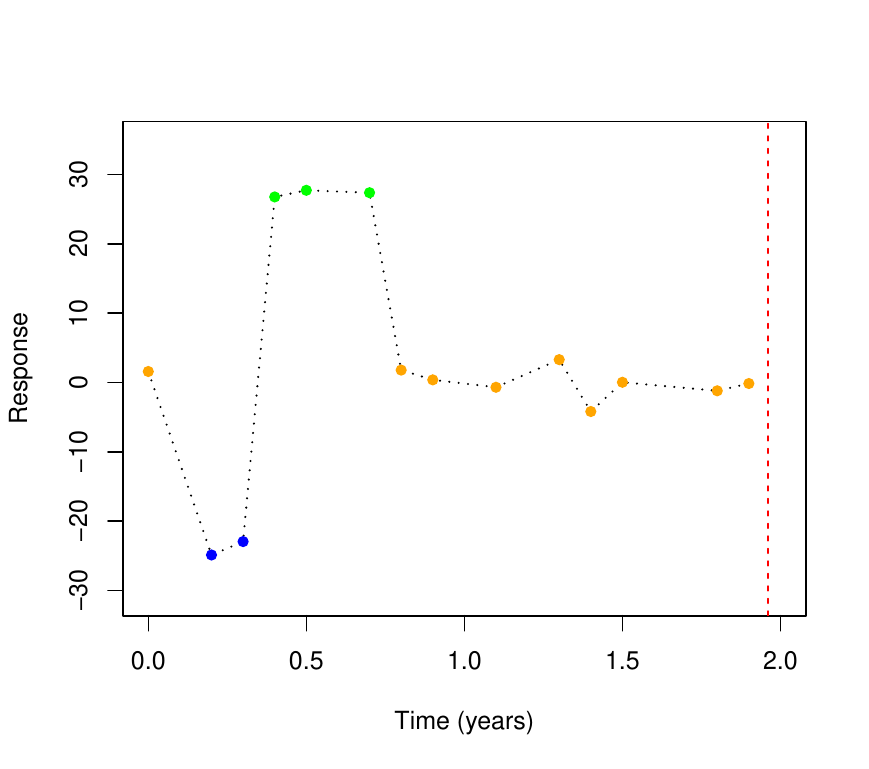}
        \caption{Subject 709}
    \end{subfigure}
    \hfill
    \begin{subfigure}{0.40\textwidth}
        \centering
        \includegraphics[width=\linewidth]{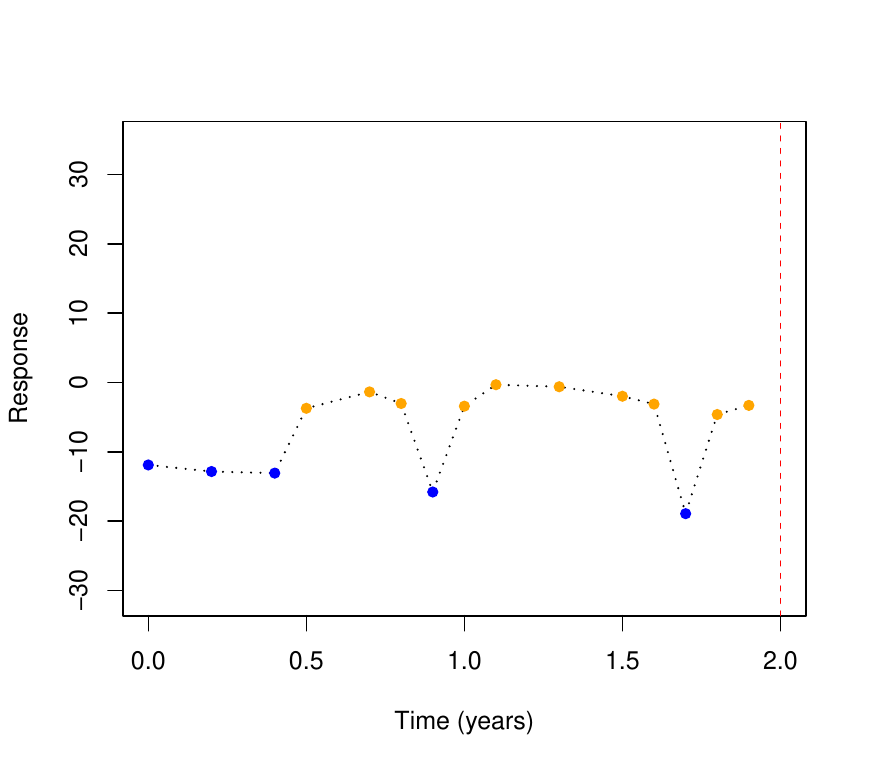}
        \caption{Subject 456}
    \end{subfigure}
    \vspace{0.5cm}
    \begin{subfigure}{0.40\textwidth}
        \centering
        \includegraphics[width=\linewidth]{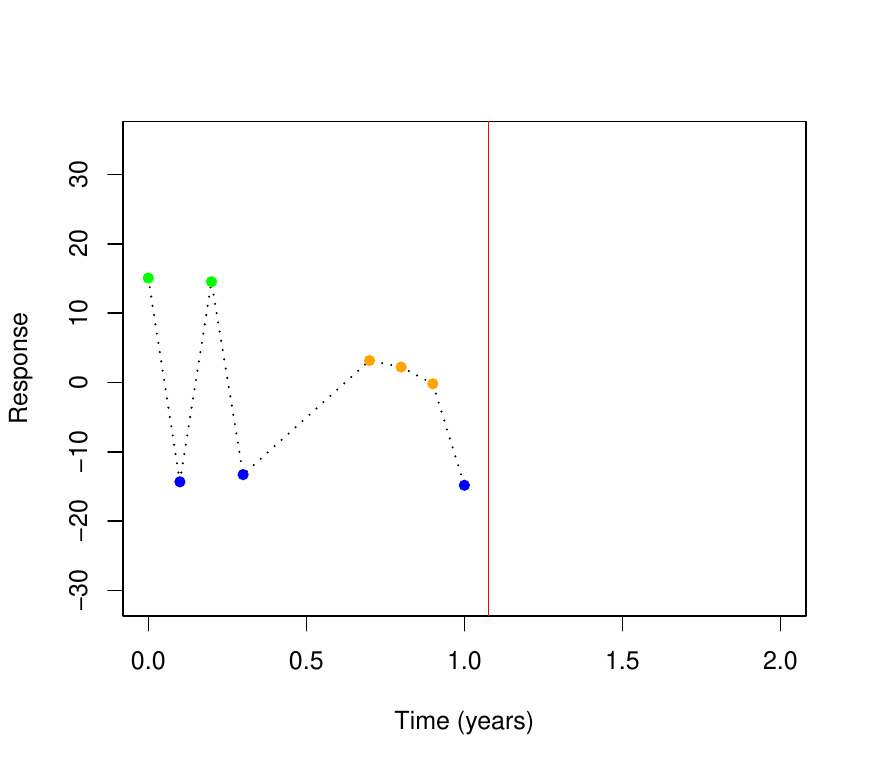}
        \caption{Subject 578}
    \end{subfigure}
\caption{Individual longitudinal trajectories for five  randomly selected subjects  for simulation study for $G=3$ classes, with observed event times indicated by a red line. Blue points represent class $1$, orange points class $2$, and green points class $3$. A solid line indicates an event, whereas a dashed line indicates censoring.}
\label{individual_trajectories_K3}  
\end{figure}

\section{Trajectories of the DJLCM with $G=3$}\label{djlcm_K3_analysis}

\setcounter{table}{0}
\renewcommand{\thetable}{B.\arabic{table}}
\setcounter{figure}{0}
\renewcommand{\thefigure}{B.\arabic{figure}}
\begin{figure}[H]
    \centering
\includegraphics[width=1\linewidth]{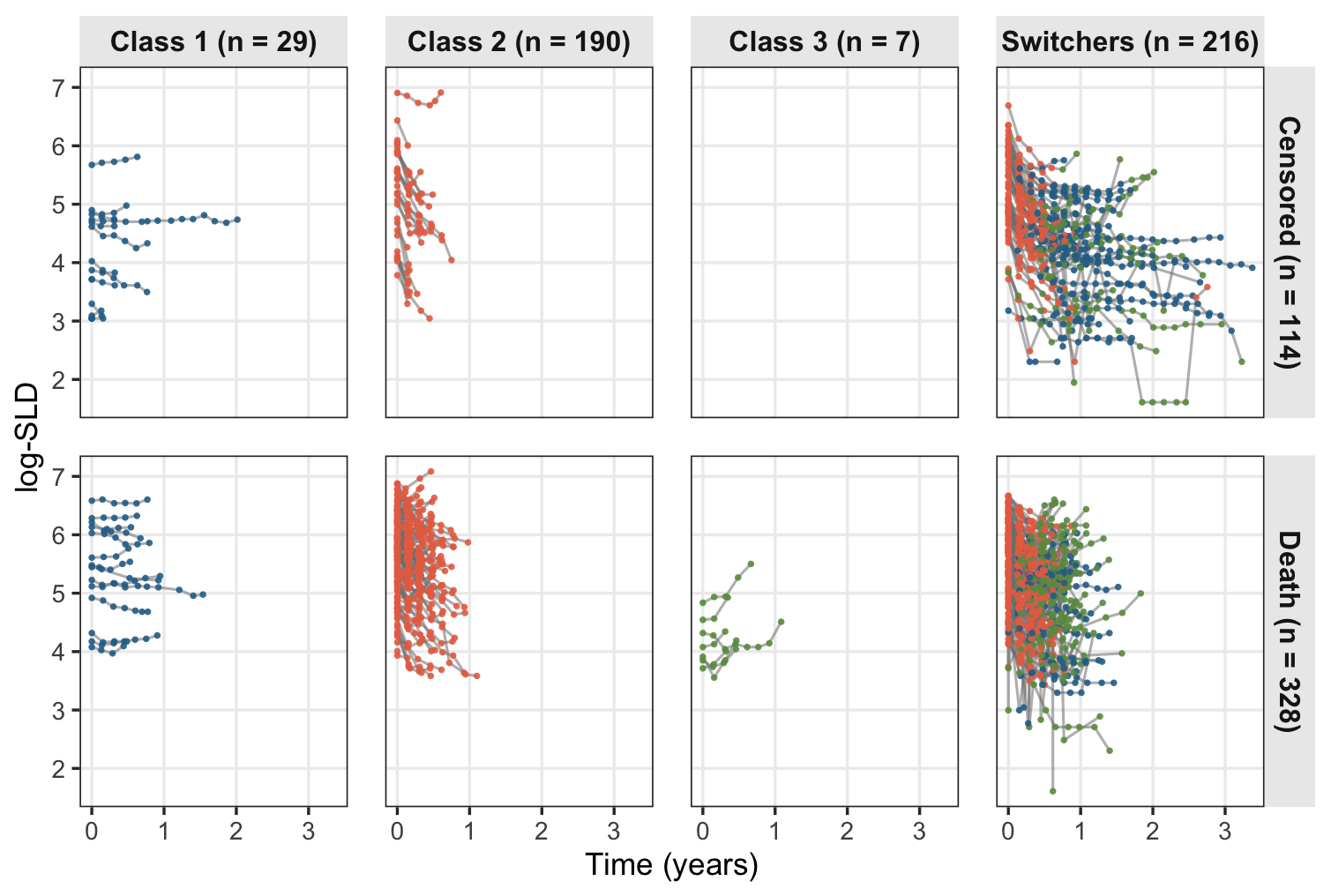}
    \caption{Longitudinal log-SLD trajectories with superimposed average trend lines, and grouped by class and event status for the DJLCM with $G=3$ for analysing the colorectal cancer data. Blue points indicate class $1$, orange points class $2$, and green points class $3$.}
    \label{traj_class_djlcm_k3}
\end{figure}

\begin{figure}[H]
\centering
\includegraphics[width=0.7\linewidth]{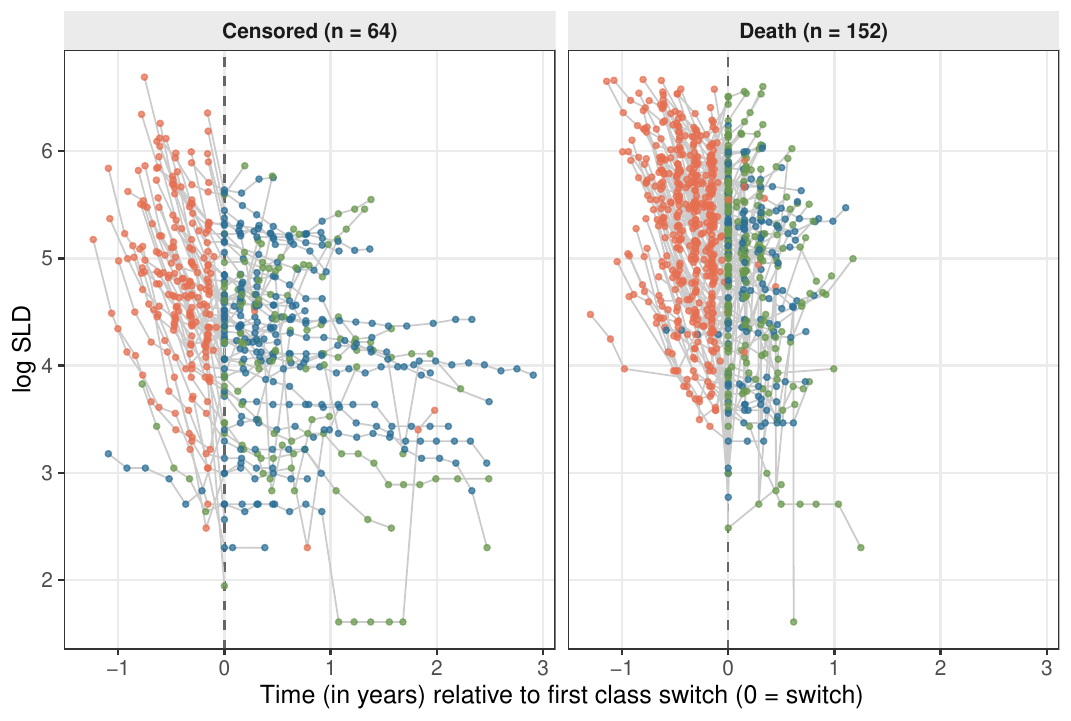}
    \caption{Longitudinal trajectory relative to first switch time for the DJLCM with $G=3$ for analysing the colorectal cancer data. Blue points indicate class $1$, orange points class $2$, and green points class $3$.}
    \label{switch}
\end{figure}

 \clearpage
 
\section{Model selection}\label{mod_selection}
\setcounter{table}{0}
\renewcommand{\thetable}{C.\arabic{table}}
\setcounter{figure}{0}
\renewcommand{\thefigure}{C.\arabic{figure}}

In this section we present the model specifications of the BJLCM and BJLCM--SRE for analysing the colorectal cancer data. Additionally, we compare these models to our DJLCM using WAIC, LPML and BIC selection criteria.

\subsection{Bayesian standard joint latent class model (BJLCM)}
The standard JLCM \cite{lin2002latent, proust2009joint} consists of the following submodel specifications.
\paragraph{Class membership submodel}
The probability of subject $i$ belonging in class $g$, $g=1,\ldots, G$, is modelled as a multinomial logistic regression model:
\begin{equation*}\label{init_cls_jlcm}
P(C_i=g)=\frac{\exp(\delta_{g0}+\delta_{g1}\text{SEX}_i)}{\sum_{h=1}^{G}\exp(\delta_{h0}+\delta_{h1}\text{SEX}_i)},\quad g=1,\ldots,G,
\end{equation*}
where   class $G$ is the reference class, with $\delta_{G0}=\delta_{G1}=0$.

\paragraph{Longitudinal submodel}
The longitudinal submodel is specified as
\begin{equation*}
\log(\text{SLD}_i(s_{ij})) \,\Big|\, _{C_i = g}
=\beta_{g0}+\beta_{g1}s_{ij}+\beta_{g2}\text{TRT}_i+\beta_{g3}\text{KRAS}_i+ b_{ig0}+b_{ig1}s_{ij}+\epsilon_{ijg},
\end{equation*}
where the vector of random effects $\bm{b}_{ig}=(b_{ig0}, b_{ig1})^\mathsf{T}$ and residual error $\epsilon_{ijg}$ are specified as in the proposed DJLCM.

\paragraph{Survival submodel}

For the survival outcome, conditional on the latent class $C_i=g$,  the hazard of the event for subject $i$ at time $t$ is modelled as

\begin{equation*}
\lambda_i(t \mid C_i=g)
= \lambda_{0g}(t) \exp\big(\gamma_{g0}+\gamma_{g1}\text{TRT}_i+\gamma_{g2}\text{KRAS}_i).
\end{equation*}
Here, $\lambda_{0g}(t)=\omega_gt^{\omega_g-1}$ denotes the class-specific
baseline hazard under the specified Weibull parameterisation.\\
The posterior class membership probabilities can be computed using the following formula \cite{proust2014joint} 

\begin{equation*}
P(C_i = g \mid \bm{\mathcal{Y}}_i, T_i, \Delta_i, \widehat{\bm{\Theta}})= \frac{P(C_i=g \mid \widehat{\bm{\Theta}})f(\bm{y}_i \mid C_i=g; \widehat{\bm{\Theta}})\lambda(T_i \mid C_i=g; \widehat{\bm{\Theta}})^{\Delta_i}S(T_i \mid C_i=g; \widehat{\bm{\Theta}})}{\sum_{g=1}^{G}P(C_i=g \mid \widehat{\bm{\Theta}})f(\bm{y}_i \mid C_i=g; \widehat{\bm{\Theta}})\lambda(T_i \mid C_i=g; \widehat{\bm{\Theta}})^{\Delta_i}S(T_i \mid C_i=g; \widehat{\bm{\Theta}})}.    
\end{equation*}
The model was fitted in a Bayesian framework using HMC.

\subsection{Bayesian JLCM with shared random effects (BJLCM--SRE)}
Following Liu et al.~\cite{liu2015joint}, we consider another JLCM with shared random effects, but implement it within a Bayesian framework using HMC, in contrast to Liu et al.~\cite{liu2015joint}, who developed the model within a frequentist framework. The class-membership and longitudinal submodels are identical to those specified for the BJLCM. The key difference is the inclusion of the subject-specific random effects in the survival submodel.

\paragraph{Survival submodel}
For the survival outcome, conditional on the latent class $C_i=g$ and the
subject-specific random effects $\bm{b}_{ig}=(b_{ig0}, b_{ig1})^\top$, the hazard of the event for
subject $i$ at time $t$ is modelled as
\begin{equation*}
\lambda_i(t\mid C_i=g,\bm{b}_{ig})
=
\lambda_{0g}(t)
\exp\left\{
\gamma_{g0}
+\gamma_{g1}\text{TRT}_i
+\gamma_{g2}\text{KRAS}_i
+\alpha_{g0} b_{ig0} +\alpha_{g1} b_{ig1}
\right\},
\end{equation*}
where $\lambda_{0g}(t)$ is specified as in the BJLCM, and $\bm{\alpha}_g=(\alpha_{g0}, \alpha_{g1})^\top$
is the class-specific association parameter vector linking the longitudinal and
survival processes through the shared random effects $\bm{b}_{ig}$. Here, the posterior class membership probabilities are calculated as in Liu et al.~\cite{liu2015joint}: 

\begin{equation*}
P(C_i = g \mid \bm{\mathcal{Y}}_i, T_i, \Delta_i, \widehat{\bm{\Theta}})= \frac{P(C_i=g \mid \widehat{\bm{\Theta}})\int f_g(\bm{y}_i \mid \bm{b})\lambda_g(T_i \mid \bm{b})^{\Delta_i}S_g(T_i \mid \bm{b})f_g(\bm{b})\,d\bm{b}}{\sum_{h=1}^{G}P(C_i=h \mid \widehat{\bm{\Theta}})\int f_h(\bm{y}_i \mid \bm{b})\lambda_h(T_i \mid \bm{b})^{\Delta_i}S_h(T_i \mid \bm{b})f_h(\bm{b})\,d\bm{b}},
\end{equation*}
where $f_g(\cdot) = f(\cdot \mid C_i = g; \widehat{\bm{\Theta}})$, and similarly for $\lambda_g$ and $S_g$.

\subsection{Model comparison criteria}\label{mod_comp}
Model selection was based on the Watanabe–Akaike Information Criterion (WAIC) \cite{watanabe2010waic}, Log Pseudo-Marginal Likelihood (LPML) \cite{gelfand1994cpo}, and Bayesian Information Criterion (BIC) \cite{schwarz1978estimating}. Let $\mathcal{D}$ denote the observed longitudinal and survival data, where $\mathcal{D}_i$ denotes the observed data for subject $i$ and $\mathcal{D}_{-i}$ denotes the data from all subjects except subject $i$. The WAIC is defined as $\mathrm{WAIC} = -2(\mathrm{lppd} - p_{\mathrm{WAIC}})$, where $\mathrm{lppd} = \sum_{i=1}^{N}\log E_{\bm{\Theta}\mid\mathcal{D}}[p(\mathcal{D}_i\mid\bm{\Theta})]$ is the log pointwise predictive density and $p_{\mathrm{WAIC}}=\sum_{i=1}^{N} \operatorname{Var}_{\text{post}}\big[\log p(\mathcal{D}_i\mid\bm{\Theta})\big]$ is the effective number of parameters, estimated by summing over the posterior variance of the each subject's log-likelihood contribution. 
The WAIC can be approximated using the formula
\begin{equation*}
\text{WAIC}=-2 \sum_{i=1}^{N}\log (\frac{1}{M}\sum_{m=1}^{M}f(\bm{y}_i, T_i, \Delta_i \mid \bm{\Theta}^{(m)})+2\sum_{i=1}^{N}\text{var}_m\left [ \log f(\bm{y}_i, T_i, \Delta_i \mid \bm{\Theta}^{(m)})\right ],
\end{equation*}
where $\bm{\Theta}_i^{(m)}$ denotes the $m$-th 
posterior HMC draw of the parameters for subject $i$, with a total of $M$ HMC draws.
\\
The LPML is defined as $\mathrm{LPML} = \sum_{i=1}^{n} \log(\mathrm{CPO}_i)$, where $\mathrm{CPO}_i = p(\mathcal{D}_i \mid \mathcal{D}_{-i})$ is the conditional predictive ordinate for subject $i$. 
It can be approximated via
\begin{equation*}
\text{LPML}=\sum_{i=1}^{N}\log\left ( \frac{1}{M} \sum_{m=1}^{M}\frac{1}{f(\bm{y}_i, T_i, \Delta_i \mid \bm{\Theta}^{(m)})}\right )^{-1}.
\end{equation*}
Lower values of WAIC indicate better model fit, whereas higher LPML values indicate better predictive performance. Also, lower BIC values are preferred.

\noindent For the Bayesian implementation, we ran $80{,}000$ HMC iterations for each model, including $40{,}000$ iterations as warm-up. The remaining $40{,}000$ iterations were used for posterior inference.

\noindent For comparison, the BJLCM was also fitted using the frequentist \texttt{lcmm} package \cite{proust2017estimation}, for which BIC is used as the model-selection criterion. The BIC values for $G=1$, $2$, and $3$ were $4027.01$, $3971.79$, and $4026.61$, respectively, with the $G=3$ model failing to converge.


\noindent The posterior summary statistics for the BJLCM ($G=3$), BJLCM--SRE ($G=3)$, and DJLCM ($G=2)$ can be found in Table \ref{posterior_model3}, Table \ref{posterior_model3_alpha}, respectively. 
 For a discussion on the comparison between the two-class DJLCM and the three-class DJLCM, the reader is referred to Section \ref{djlcm_est_traj}.


\begin{table}[htbp]
\centering
\caption{Model selection criteria for the BJLCM ($G=1,2,3,4)$, BJLCM--SRE with $G=1,2,3$ and the DJLCM with $G=2,3$ latent classes.}
\label{model-comparison}
\resizebox{\textwidth}{!}{%
\begin{tabular}{l rrrr rrr rr}
\toprule
 & \multicolumn{4}{c}{BJLCM} & \multicolumn{3}{c}{BJLCM--SRE} & \multicolumn{2}{c}{DJLCM} \\
\cmidrule(lr){2-5} \cmidrule(lr){6-8} \cmidrule(lr){9-10}
 & $G=1$ & $G=2$ & $G=3$ & $G=4$ & $G=1$ & $G=2$ & $G=3$ & $G=2$ & $G=3$ \\
\midrule
LPML & $-1995.368$ & $-1718.659$ & $-1672.568$ & $-1640.842$
     & $-1942.380$ & $-1684.780$ & $-1647.941$
     & $-1425.500$ & $-1275.379$ \\
WAIC & $3990.653$ & $3436.876$ & $3339.450$ & $3276.411$
     & $3884.846$ & $3369.364$ & $3301.897$
     & $2851.402$ & $2533.928$ \\
BIC  & $4027.089$ & $3538.436$ & $3476.188$ & $3470.668$
     & $3930.143$ & $3485.151$ & $3452.134$
     & $2969.302$ & $2740.201$ \\
\bottomrule
\end{tabular}%
}
\end{table}

\begin{table}[htbp]
\centering
\small
\setlength{\tabcolsep}{4pt}
\caption{Parameter estimates (Mean), standard deviation (SD), $2.5\%$ lower bound of $95\%$ credible interval, $97.5\%$ upper bound of $95\%$ credible interval and $\widehat{R}$: Gelman-Rubin diagnostic for analysing the Colorectal Cancer data using the DJLCM with $G=3$.}
\label{posterior_djlcm_g3}
\resizebox{\textwidth}{!}{%
\begin{tabular}{l@{\ }l rrrrr c rrrrr c rrrrr}
\toprule
 & & \multicolumn{5}{c}{Class 1 ($g=1$)} & & \multicolumn{5}{c}{Class 2 ($g=2$)} & & \multicolumn{5}{c}{Class 3 ($g=3$)} \\
\cmidrule(lr){3-7} \cmidrule(lr){9-13} \cmidrule(lr){15-19}
\multicolumn{2}{l}{Parameter} & Mean & SD & 2.5\% & 97.5\% & $\hat{R}$ & &
Mean & SD & 2.5\% & 97.5\% & $\hat{R}$ & &
Mean & SD & 2.5\% & 97.5\% & $\hat{R}$ \\
\midrule
\multicolumn{19}{l}{\textit{Initial class membership submodel} } \\
Intercept     & ($\delta_{g0}$)  & -2.32 & 0.33 & -3.02 & -1.73 & 1.00 & & \multicolumn{5}{c}{Reference class} & & -3.04 & 0.71 & -4.59 & -1.80 & 1.00 \\
\text{SEX}    & ($\delta_{g1}$)  & 0.16  & 0.54 & -0.97 & 1.15  & 1.00 & & \multicolumn{5}{c}{} & & 0.63  & 0.89 & -1.36 & 2.20  & 1.00 \\
\addlinespace
\multicolumn{19}{l}{\textit{Class transition submodel} (from class $g$)} \\
Intercept     & ($\eta_{g10}$)   & 4.27  & 0.62 & 3.20  & 5.63  & 1.00 & & -1.94 & 0.28 & -2.55 & -1.45 & 1.00 & & 0.63  & 0.72 & -0.80 & 2.02  & 1.00 \\
\text{SEX}    & ($\eta_{g11}$)   & 1.33  & 1.06 & -0.59 & 3.55  & 1.00 & & 0.14  & 0.35 & -0.54 & 0.83  & 1.00 & & 0.16  & 1.45 & -3.03 & 2.79  & 1.05 \\
Intercept     & ($\eta_{g30}$)   & 1.95  & 0.70 & 0.68  & 3.41  & 1.00 & & -1.70 & 0.25 & -2.23 & -1.26 & 1.00 & & 3.48  & 0.50 & 2.59  & 4.55  & 1.00 \\
\text{SEX}    & ($\eta_{g31}$)   & 1.02  & 1.13 & -1.10 & 3.33  & 1.00 & & -0.45 & 0.39 & -1.27 & 0.27  & 1.00 & & 0.63  & 0.98 & -1.06 & 2.79  & 1.02 \\
\addlinespace
\multicolumn{19}{l}{\textit{Longitudinal submodel}} \\
Intercept     & ($\beta_{g0}$)   & 4.81  & 0.18 & 4.43  & 5.15  & 1.00 & & 5.56  & 0.07 & 5.43  & 5.69  & 1.00 & & 4.59  & 0.17 & 4.25  & 4.90  & 1.00 \\
Time          & ($\beta_{g1}$)   & -0.20 & 0.05 & -0.30 & -0.11 & 1.00 & & -1.23 & 0.10 & -1.44 & -1.04 & 1.00 & & 0.27  & 0.12 & 0.05  & 0.51  & 1.00 \\
\text{TRT}    & ($\beta_{g2}$)   & -0.06 & 0.20 & -0.45 & 0.34  & 1.00 & & -0.03 & 0.08 & -0.18 & 0.12  & 1.00 & & -0.02 & 0.18 & -0.37 & 0.32  & 1.00 \\
\text{KRAS}   & ($\beta_{g3}$)   & 0.06  & 0.21 & -0.36 & 0.47  & 1.01 & & -0.03 & 0.08 & -0.17 & 0.12  & 1.00 & & 0.16  & 0.18 & -0.19 & 0.52  & 1.00 \\
$\sigma_{g}$  &                  & 0.05  & $<$ 0.01 & 0.04 & 0.06 & 1.00 & & 0.13 & 0.01 & 0.11 & 0.14 & 1.00 & & 0.16 & 0.01 & 0.14 & 0.19 & 1.00 \\
$D_{g,11}$    &                  & 0.90  & 0.13 & 0.67  & 1.19  & 1.00 & & 0.46  & 0.04 & 0.38  & 0.55  & 1.00 & & 0.55  & 0.12 & 0.34  & 0.81  & 1.00 \\
$D_{g,12}$    &                  & -0.02 & 0.04 & -0.11 & 0.05  & 1.00 & & 0.20  & 0.07 & 0.07  & 0.33  & 1.00 & & 0.12  & 0.09 & -0.07 & 0.28  & 1.00 \\
$D_{g,22}$    &                  & 0.08  & 0.03 & 0.04  & 0.14  & 1.00 & & 1.08  & 0.19 & 0.72  & 1.48  & 1.00 & & 0.36  & 0.11 & 0.19  & 0.61  & 1.00 \\
\addlinespace
\multicolumn{19}{l}{\textit{Survival submodel}} \\
$\omega_{g}$  &                  & 2.36  & 0.44 & 1.58  & 3.31  & 1.00 & & 1.79  & 0.20 & 1.46  & 2.24  & 1.00 & & 1.82  & 0.36 & 1.16  & 2.59  & 1.00 \\
Intercept     & ($\gamma_{g0}$)  & -3.16 & 0.67 & -4.62 & -1.98 & 1.00 & & -1.41 & 0.30 & -2.07 & -0.90 & 1.00 & & -1.55 & 0.47 & -2.51 & -0.66 & 1.00 \\
\text{TRT}    & ($\gamma_{g1}$)  & 0.19  & 0.44 & -0.69 & 1.05  & 1.00 & & 0.40  & 0.26 & -0.10 & 0.93  & 1.00 & & -0.05 & 0.28 & -0.61 & 0.49  & 1.00 \\
\text{KRAS}   & ($\gamma_{g2}$)  & 0.61  & 0.44 & -0.24 & 1.50  & 1.00 & & 0.27  & 0.26 & -0.25 & 0.79  & 1.00 & & 0.31  & 0.29 & -0.25 & 0.89  & 1.00 \\
$\alpha_{g0}$ &                  & 1.21  & 0.27 & 0.71  & 1.79  & 1.00 & & $<-0.00$ & 0.28 & -0.60 & 0.50 & 1.00 & & 0.73  & 0.32 & 0.14  & 1.39  & 1.00 \\
$\alpha_{g1}$ &                  & 0.32  & 0.83 & -1.33 & 1.94  & 1.00 & & 1.51  & 0.33 & 0.99  & 2.28  & 1.00 & & 0.65  & 0.46 & -0.24 & 1.56  & 1.00 \\
\bottomrule
\end{tabular}%
}
\end{table}
     



\begin{table}[htbp]
\centering
\footnotesize
\setlength{\tabcolsep}{2.5pt}
\caption{
Parameter estimates (Mean), standard deviation (SD), $2.5\%$ lower bound of $95\%$ credible interval, $97.5\%$ upper bound of $95\%$ credible interval and $\widehat{R}$: Gelman-Rubin diagnostic for analysing the Colorectal Cancer data using the BJLCM with $G=3$.}
\label{posterior_model3}
\resizebox{\textwidth}{!}{%
\begin{tabular}{@{}l@{\ }l rrrrr c rrrrr c rrrrr@{}}
\toprule
 & & \multicolumn{5}{c}{Class 1 ($g=1$)} & & \multicolumn{5}{c}{Class 2 ($g=2$)} & & \multicolumn{5}{c}{Class 3 ($g=3$)} \\
\cmidrule(lr){3-7} \cmidrule(lr){9-13} \cmidrule(lr){15-19}
\multicolumn{2}{@{}l}{Parameter} & Mean & SD & 2.5\% & 97.5\% & $\hat{R}$ & &
Mean & SD & 2.5\% & 97.5\% & $\hat{R}$ & &
Mean & SD & 2.5\% & 97.5\% & $\hat{R}$ \\
\midrule
\multicolumn{19}{@{}l}{\textit{Class membership submodel}} \\
Intercept     & ($\delta_{g0}$) & 0.55  & 0.46 & -0.44 & 1.42  & 1.01 & & 1.78  & 0.25 & 1.31  & 2.29  & 1.00 & & \multicolumn{5}{c}{Reference class} \\
\text{SEX}    & ($\delta_{g1}$) & -0.02 & 0.47 & -0.94 & 0.91  & 1.00 & & -0.07 & 0.39 & -0.81 & 0.73  & 1.00 & & \multicolumn{5}{c}{} \\
\addlinespace
\multicolumn{19}{@{}l}{\textit{Longitudinal submodel}} \\
Intercept     & ($\beta_{g0}$)  & 5.59  & 0.24 & 4.94  & 5.92  & 1.01 & & 5.31  & 0.12 & 5.06  & 5.53  & 1.00 & & 4.78  & 0.27 & 4.28  & 5.36  & 1.00 \\
Time          & ($\beta_{g1}$)  & -0.14 & 0.10 & -0.36 & 0.04  & 1.00 & & -0.65 & 0.04 & -0.73 & -0.57 & 1.01 & & -0.79 & 0.16 & -1.12 & -0.48 & 1.00 \\
\text{TRT}    & ($\beta_{g2}$)  & 0.48  & 0.24 & 0.06  & 0.99  & 1.00 & & -0.12 & 0.11 & -0.32 & 0.09  & 1.00 & & -0.12 & 0.27 & -0.68 & 0.39  & 1.00 \\
\text{KRAS}   & ($\beta_{g3}$)  & -0.42 & 0.24 & -0.87 & 0.10  & 1.01 & & 0.21  & 0.11 & -0.01 & 0.43  & 1.01 & & -0.29 & 0.26 & -0.81 & 0.21  & 1.00 \\
$\sigma_{g}$  &                 & 0.09  & 0.02 & 0.04  & 0.11  & 1.02 & & 0.24  & 0.01 & 0.22  & 0.26  & 1.01 & & 0.61  & 0.04 & 0.54  & 0.70  & 1.00 \\
$D_{g,11}$    &                 & 0.36  & 0.12 & 0.20  & 0.67  & 1.00 & & 0.49  & 0.06 & 0.39  & 0.62  & 1.00 & & 0.66  & 0.19 & 0.37  & 1.10  & 1.00 \\
$D_{g,12}$    &                 & -0.03 & 0.05 & -0.13 & 0.08  & 1.00 & & 0.01  & 0.03 & -0.05 & 0.06  & 1.00 & & -0.25 & 0.16 & -0.62 & -0.01 & 1.00 \\
$D_{g,22}$    &                 & 0.23  & 0.14 & 0.04  & 0.55  & 1.00 & & 0.13  & 0.03 & 0.07  & 0.20  & 1.00 & & 0.42  & 0.24 & 0.08  & 0.99  & 1.00 \\
\addlinespace
\multicolumn{19}{@{}l}{\textit{Survival submodel}} \\
$\omega_{g}$  &                 & 1.34  & 0.18 & 1.04  & 1.75  & 1.00 & & 1.34  & 0.10 & 1.16  & 1.56  & 1.00 & & 1.89  & 0.41 & 1.14  & 2.75  & 1.00 \\
Intercept     & ($\gamma_{g0}$) & -0.23 & 0.34 & -1.03 & 0.37  & 1.00 & & -1.29 & 0.21 & -1.76 & -0.92 & 1.00 & & -2.59 & 0.62 & -3.84 & -1.41 & 1.00 \\
\text{TRT}    & ($\gamma_{g1}$) & 0.43  & 0.34 & -0.22 & 1.12  & 1.00 & & 0.12  & 0.16 & -0.18 & 0.43  & 1.00 & & -0.41 & 0.51 & -1.41 & 0.61  & 1.00 \\
\text{KRAS}   & ($\gamma_{g2}$) & -0.71 & 0.45 & -1.68 & 0.12  & 1.01 & & 0.56  & 0.17 & 0.22  & 0.89  & 1.00 & & 0.30  & 0.51 & -0.74 & 1.29  & 1.00 \\
\bottomrule
\end{tabular}%
}
\end{table}
\FloatBarrier

\begin{table}[htbp]
\centering
\footnotesize
\setlength{\tabcolsep}{2.5pt}
\caption{
Parameter estimates (Mean), standard deviation (SD), $2.5\%$ lower bound of $95\%$ credible interval, $97.5\%$ upper bound of $95\%$ credible interval and $\widehat{R}$: Gelman-Rubin diagnostic for analysing the Colorectal Cancer data using the BJLCM--SRE with $G=3$.}
\label{posterior_model3_alpha}
\resizebox{\textwidth}{!}{%
\begin{tabular}{@{}l@{\ }l rrrrr c rrrrr c rrrrr@{}}
\toprule
 & & \multicolumn{5}{c}{Class 1 ($g=1$)} & & \multicolumn{5}{c}{Class 2 ($g=2$)} & & \multicolumn{5}{c}{Class 3 ($g=3$)} \\
\cmidrule(lr){3-7} \cmidrule(lr){9-13} \cmidrule(lr){15-19}
\multicolumn{2}{@{}l}{Parameter} & Mean & SD & 2.5\% & 97.5\% & $\hat{R}$ & &
Mean & SD & 2.5\% & 97.5\% & $\hat{R}$ & &
Mean & SD & 2.5\% & 97.5\% & $\hat{R}$ \\
\midrule
\multicolumn{19}{@{}l}{\textit{Class membership submodel}} \\
Intercept     & ($\delta_{g0}$) & -2.01 & 0.52 & -2.92 & -0.93 & 1.00 & & \multicolumn{5}{c}{Reference class} & & -1.92 & 0.25 & -2.44 & -1.44 & 1.00 \\
\text{SEX}    & ($\delta_{g1}$) & -0.06 & 0.51 & -1.10 & 0.87  & 1.00 & & \multicolumn{5}{c}{} & & -0.01 & 0.40 & -0.83 & 0.75  & 1.00 \\
\addlinespace
\multicolumn{19}{@{}l}{\textit{Longitudinal submodel}} \\
Intercept     & ($\beta_{g0}$)  & 5.13  & 0.38 & 4.33  & 5.79  & 1.00 & & 5.39  & 0.09 & 5.20  & 5.54  & 1.00 & & 4.89  & 0.27 & 4.37  & 5.44  & 1.00 \\
Time          & ($\beta_{g1}$)  & -0.10 & 0.10 & -0.28 & 0.12  & 1.00 & & -0.59 & 0.05 & -0.69 & -0.50 & 1.00 & & -0.74 & 0.16 & -1.06 & -0.42 & 1.00 \\
\text{TRT}    & ($\beta_{g2}$)  & 0.56  & 0.38 & -0.15 & 1.34  & 1.00 & & -0.06 & 0.10 & -0.25 & 0.12  & 1.00 & & -0.15 & 0.29 & -0.74 & 0.41  & 1.00 \\
\text{KRAS}   & ($\beta_{g3}$)  & -0.02 & 0.36 & -0.67 & 0.74  & 1.00 & & 0.09  & 0.10 & -0.10 & 0.29  & 1.00 & & -0.32 & 0.28 & -0.87 & 0.24  & 1.00 \\
$\sigma_{g}$  &                 & 0.05  & 0.02 & 0.03  & 0.09  & 1.00 & & 0.23  & 0.01 & 0.21  & 0.25  & 1.00 & & 0.62  & 0.04 & 0.54  & 0.71  & 1.00 \\
$D_{g,11}$    &                 & 0.63  & 0.25 & 0.28  & 1.24  & 1.00 & & 0.51  & 0.05 & 0.41  & 0.63  & 1.00 & & 0.67  & 0.20 & 0.36  & 1.12  & 1.00 \\
$D_{g,12}$    &                 & -0.01 & 0.07 & -0.15 & 0.15  & 1.00 & & 0.02  & 0.03 & -0.03 & 0.08  & 1.00 & & -0.19 & 0.15 & -0.54 & 0.04  & 1.00 \\
$D_{g,22}$    &                 & 0.14  & 0.14 & 0.02  & 0.54  & 1.00 & & 0.17  & 0.04 & 0.10  & 0.24  & 1.00 & & 0.40  & 0.24 & 0.06  & 0.97  & 1.00 \\
\addlinespace
\multicolumn{19}{@{}l}{\textit{Survival submodel}} \\
$\omega_{g}$  &                 & 1.60  & 0.36 & 0.98  & 2.37  & 1.00 & & 1.41  & 0.09 & 1.25  & 1.60  & 1.00 & & 2.03  & 0.46 & 1.19  & 2.97  & 1.00 \\
Intercept     & ($\gamma_{g0}$) & -0.66 & 0.68 & -2.12 & 0.52  & 1.00 & & -1.25 & 0.19 & -1.68 & -0.94 & 1.00 & & -2.66 & 0.71 & -4.10 & -1.33 & 1.00 \\
\text{TRT}    & ($\gamma_{g1}$) & 0.09  & 0.61 & -1.22 & 1.25  & 1.00 & & 0.28  & 0.16 & -0.03 & 0.59  & 1.00 & & -0.66 & 0.60 & -1.87 & 0.52  & 1.00 \\
\text{KRAS}   & ($\gamma_{g2}$) & -0.21 & 0.72 & -1.63 & 1.18  & 1.00 & & 0.42  & 0.19 & 0.08  & 0.83  & 1.00 & & 0.57  & 0.60 & -0.62 & 1.75  & 1.00 \\
$\alpha_{g0}$ &                 & 0.85  & 0.45 & 0.04  & 1.81  & 1.00 & & 0.74  & 0.12 & 0.51  & 0.99  & 1.00 & & 1.01  & 0.49 & 0.08  & 2.00  & 1.00 \\
$\alpha_{g1}$ &                 & 0.56  & 1.26 & -1.91 & 2.85  & 1.00 & & 0.53  & 0.33 & -0.19 & 1.11  & 1.00 & & 0.92  & 0.78 & -0.71 & 2.44  & 1.00 \\
\bottomrule
\end{tabular}%
}
\end{table}

\FloatBarrier

\begin{figure}[htbp]
\centering
\begin{subfigure}[t]{0.5\textwidth}
\includegraphics[width=\linewidth]{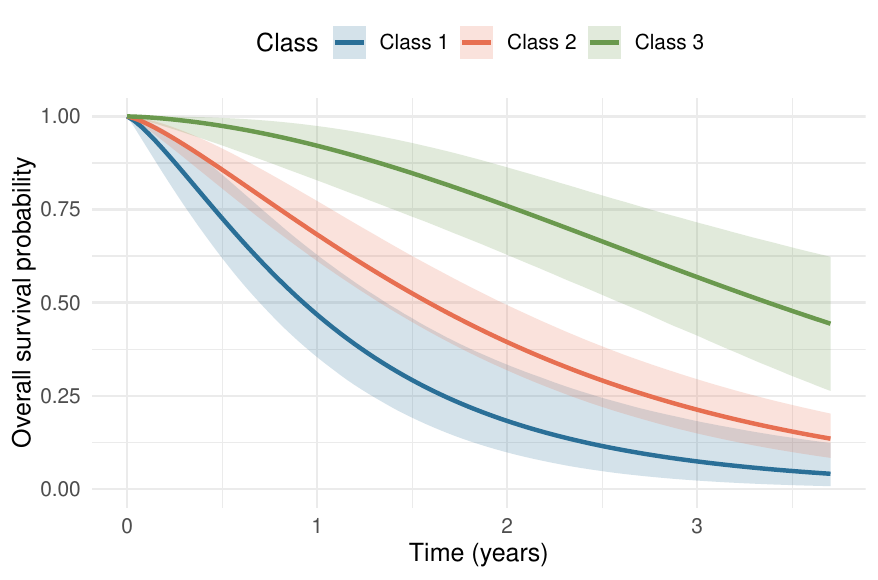}
  \caption{BJLCM (G=3)}
  \label{fig:a_surv}
\end{subfigure}\hfill
\begin{subfigure}[t]{0.5\textwidth}
  \includegraphics[width=\linewidth]{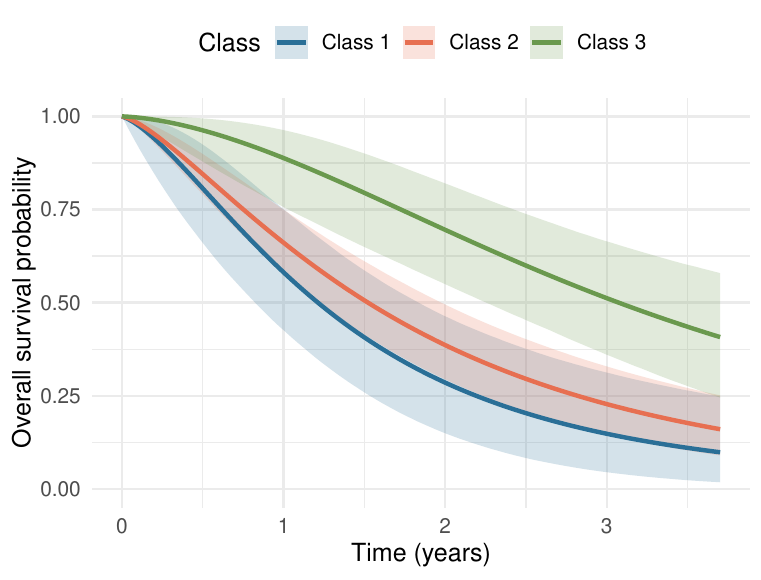}
  \caption{BJLCM--SRE (G=3)}
  \label{fig:b_surv}
\end{subfigure}
\caption{Survival curves for the BJLCM with $G=3$ (panel a) and the BJLCM--SRE with $G=3$ (panel b) for analysing the colorectal cancer data.}
\label{fig:4_models_surv}
\end{figure}

\end{document}